\documentclass[pr,twocolumn,showpacs]{revtex4}
\usepackage{bm}
\usepackage{graphicx}
\usepackage{amsmath}
\usepackage{eufrak}
\usepackage{color}
\usepackage{upgreek}
\usepackage{subfigure}
\usepackage[hidelinks]{hyperref}

\usepackage{placeins}
\usepackage{lipsum}

\newcommand{\nix}[1]{}

\begin{document}

\title{Magnetic quantum ratchet effect in
(Cd,Mn)Te- and CdTe-based quantum well structures
with a lateral asymmetric superlattice}

\author{P.\,Faltermeier,$^1$  G.V. Budkin,$^2$ J.\,Unverzagt,$^1$ S.\,Hubmann,$^1$ A.\,Pfaller,$^1$
V.V.\,Bel'kov,$^2$  L.E. Golub,$^2$ E.L. Ivchenko,$^2$  Z. Adamus,$^3$ G. Karczewski,$^3$ T. Wojtowicz,$^{3,4}$
V.V.\,Popov,$^{5,6}$ D.V.\,Fateev,$^{5}$ D.A. Kozlov,$^{7}$
D.\,Weiss,$^1$ and S.D.\,Ganichev$^1$}

\affiliation{$^1$Terahertz Center, University of Regensburg, 93040 Regensburg, Germany}

\affiliation{$^2$Ioffe Institute, 194021 St.\,Petersburg, Russia}

\affiliation{$^3$ Institute of  Physics, Polish Academy of Sciences, al. Lotnik\'{o}w 32/46, PL 02-668 Warszawa, Poland}

\affiliation{$^4$ International Research Centre MagTop, al. Lotnik\'{o}w 32/46, PL 02-668 Warszawa, Poland}

\affiliation{$^5$ Institute of Radio Engineering and Electronics (Saratov Branch), 410019 Saratov, Russia}

\affiliation{$^6$ Saratov State University, 410012 Saratov, Russia}

\affiliation{$^7$ A.V. Rzhanov Institute of Semiconductor Physics, 630090 Novosibirsk,  Russia}

\begin{abstract}
We report on the observation of magnetic quantum ratchet effect in
(Cd,Mn)Te- and CdTe-based quantum well structures with an
asymmetric lateral dual grating gate superlattice subjected to an
external magnetic field applied normal to the quantum well plane. A  $dc$
electric current excited by \textit{cw} terahertz laser radiation
shows $1/B$-oscillations with an amplitude much larger as compared to the photocurrent at zero
magnetic field. We show that the photocurrent  is caused by the
combined action of a spatially periodic in-plane potential and the
spatially modulated radiation due to the near field effects of 
light diffraction. Magnitude and direction of the photocurrent
are determined by the degree of the lateral asymmetry controlled by the variation of voltages
applied to the individual gates. The observed magneto-oscillations
with enhanced photocurrent amplitude 
%are shown to be also caused by the magnetic quantum ratchet effect. The photocurrent oscillations
result from Landau quantization and, for (Cd,Mn)Te at low
temperatures, from the exchange enhanced Zeeman splitting in diluted
magnetic heterostructures. Theoretical analysis, considering the
magnetic quantum ratchet effect in the framework of semiclassical
approach, describes quite well the experimental results.
%including, magneto-oscillations, photocurrent enhancement,
%its variation upon change of voltages applied to gates,
%as well as temperature  dependencies.
\end{abstract}

\maketitle

\section{Introduction}

%\sdg{***SdH ***power ***alpha, Calibration, heating with Vasya, Popov's calculations, Fig, Lenya's text}

Spatially periodic non-centrosymmetric systems are able to
transport nonequilibrium particles in the absence of an average macroscopic force resulting in the ratchet effect~\cite{prost,reimann,applphys,hanggi,Denisovhanggi}.
Ratchet effects whose prerequisites are simultaneous breaking of both thermal equilibrium and spatial inversion symmetry can be realized in a great variety of forms and in particular, as electric transport in semiconductor systems~\cite{buttiker,buttiker2,grifoni,kotthaus,samuelson,Chepel,Chepelianskii,Kvon,Olbrich_PRL_09,Review_JETP_Lett,Nalitov,PopovAPL2013,Koniakhin2014,Rozhansky2015,PopovIvchenko}.
Recent experiments demonstrated that terahertz radiation induced ratchet effects, can be efficiently excited
in semiconductor quantum wells (QW)~\cite{Olbrich_PRL_09,Review_JETP_Lett,Olbrich_PRB_11,Kannan} and graphene~\cite{ratchet_graphene,Drexler13} with lateral superlattice structures.
These experiments allowed one to explore basic physics of the ratchet effects in low dimensional electron systems, provide information on non-equilibrium transport in such systems and demonstrate that ratchet effects can be applied for room temperature terahertz radiation detection~\cite{otsuji,det2,otsuji2,Popov_Otsuji_Knap}.  The latter, besides high sensitivity and short response times, offer new functionality being a good candidate for all-electric detection of the radiation polarization state including radiation helicity~\cite{Otsuji_Ganichev,helicitydetector,ellipticitydetector,ellipticitydetector2}. 

Here we report on the observation and study of
magnetic quantum ratchet effect in (Cd,Mn)Te/(Cd,Mg)Te diluted magnetic heterostructures and CdTe/CdMgTe QWs superimposed with lateral asymmetric superlattices. Applying magnetic field $B$ along the growth direction
we observe that the ratchet current exhibits sign-alternating $1/B$-periodic oscillations 
with amplitudes immensely larger than the ratchet signal at zero magnetic field.
The results are analyzed in terms of the theory of magnetic ratchet effects in QW structures 
with a lateral asymmetric periodic potential~\cite{JETP_Lett_review}. We show that the photocurrent generation is based on the combined action of a spatially periodic in-plane potential and the spatially modulated light
due to the near field effects of radiation diffraction.
Corresponding theoretical analysis describes the experimental results.

The paper is organized as follows. First, we  present our samples
and results of magneto-transport characterization
(Sec.~\ref{sample}), as well as briefly describe the experimental
technique (Sec. \ref{experi}). In Sec.~\ref{erstes} we discuss the
results on the photocurrents generated at zero magnetic
field. Section~\ref{magnetic} contains the main experimental
results on the magnetic quantum ratchet effect. It has three
subsections describing the basic results, the laser beam scan
across the structure and the effect of the electrostatic potential acting on the electron gas. In the following
Sec.~\ref{discussion} we present the theory, calculate 
dependencies of the photocurrent on magnetic field, and compare them
with the experimental data. Finally, in Sec.
\ref{summary}, we summarize the results and discuss prospectives
of future experimental studies of the magnetic quantum ratchet effect.

\section{Samples} \label{sample}
\subsection{Samples grow and characterization}
\FloatBarrier

\begin{figure}[h]
\includegraphics[width=\linewidth]{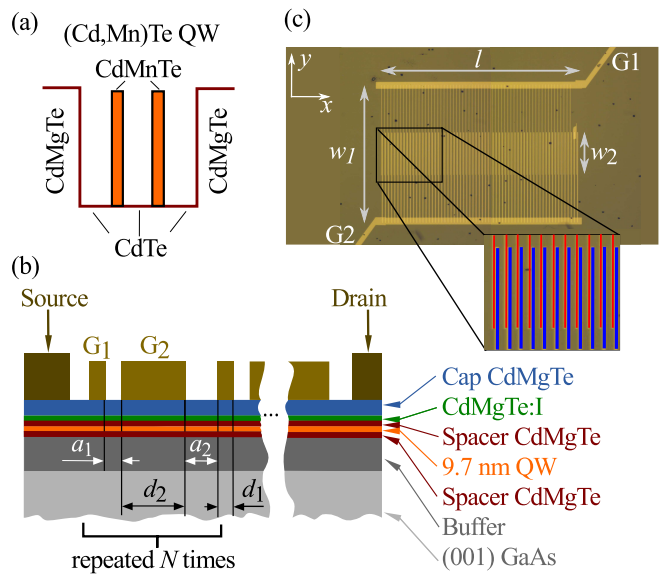}
\caption{ (a) Sketch of CdMgTe/(Cd,Mn)Te/CdMgTe QW. Two red bars
show evenly spaced three-monolayer-thick Cd$_{0.8}$Mn$_{0.2}$Te
layers inserted during QW growth. (b) Cross-section of the
dual-grating gate superlattice formed by metal fingers deposited
on top of the QW structure. Structure composition is given on the right side of the sketch. The
supercell of the grating gate fingers consists of metal stripes
having two different widths $d_1$ and $d_2$ separated by spacings
$a_1$ and $a_2$. This asymmetric supercell is  repeated $N$ times
to create a superlattice with period $d = d_1+a_1+
d_2+a_2$, see Table~\ref{tab1}. (c) Photograph of the sample together with 
schematic top view of the dual-grating gate
superlattice. Thin red and thick blue lines sketch the top gates
having different thicknesses and spacings.}
\label{structure}
\end{figure}

Experiments are carried out on (Cd,Mn)Te/CdMgTe and CdTe/CdMgTe
single QW structures grown by molecular beam epitaxy on (001)-oriented GaAs
substrates~\cite{Crooker,Egues,Jaroszynski2002,DMSPRL09,DMS2PRB12}.
During the growth the fluxes of Cd, Te, Mg and Mn have been
supplied from elemental sources while Iodine flux has been
obtained from a ZnI$_2$ source. The schematic layout of the (Cd,Mn)Te/CdMgTe layer structure and sketch of
the QW are presented in Fig.~\ref{structure}. The
thick buffer ($\approx$~6~$\upmu$m), consisting of CdTe and
Cd$_{0.76}$Mg$_{0.24}$Te layers and CdTe/Cd$_{0.76}$Mg$_{0.24}$Te
short period superlattices, have been grown to reduce the number of
dislocations resulting from the strong lattice mismatch between
GaAs and Cd$_{1-x}$Mg$_x$Te.
The well width in both kinds of structures is 9.7~nm and the
Cd$_{0.76}$Mg$_{0.24}$Te alloy serves as a barrier
material. The composition of barriers has been determined from
photoluminescence spectra. In order to obtain a two-dimensional
electron gas the structures have been modulation doped by Iodine
donors incorporated  into the 5~nm thick region of the top barrier
at the distance of either 10 or 15~nm away from the QW, in
(Cd,Mn)Te/CdMgTe and CdTe/CdMgTe structures, respectively.
Doped regions have been overgrown by an undoped cap layer with thickness of either 50~nm, for (Cd,Mn)Te/CdMgTe QWs, or 75~nm,
for  CdTe/CdMgTe QWs. The density of the two-dimensional electron gas,
$n_e$,  and the electron mobility, $\mu$, determined at liquid helium
temperature (4.2 K) for (Cd,Mn)Te/CdMgTe structures  are $n_e =
6.6 \times 10^{11}$ cm$^{-2}$
%EF=13 meV, N=6.6 10^{11} cm^{-2}
and
$\mu = 9.5 \times 10^3$~cm$^{2}$/Vs, and
for  CdTe/CdMgTe structures 
$n_e = 4.2 \times 10^{11}$~cm$^{-2}$ and
$\mu = 65 \times 10^3$~cm$^2$/Vs.

(Cd,Mn)Te/CdMgTe structures contain a single QW made
of (Cd,Mn)Te digital magnetic alloy~\cite{Kneip2006}, while
CdTe/CdMgTe structures contain a single QW made of non-magnetic
CdTe. In the former structures two evenly spaced
Cd$_{0.8}$Mn$_{0.2}$Te thin layers   were inserted during the QW
growth, see Fig.~\ref{structure}(a). These two layers were
three-monolayer-thick (ML) and separated from each other
and from the barriers by 8 ML thick CdTe (1 ML is 0.324 nm thick).

Incorporation  of Mn$^{2+}$ ions  carrying
localized spin $S = 5/2$ into the QW region leads to a strong enhancement of the
effective $g$ factor of band carriers, and hence to an enhanced
Zeeman splitting. This has been shown in previous
magneto-photoluminescence (PL) studies performed on the
(Cd,Mn)Te/CdMgTe samples made from the same wafer as used in
the current study. Magneto-PL showed strong red-shift of the PL line
with increasing magnetic field~\cite{DMSPRL09}. From fitting
of the modified Brillouin function~\cite{Gaj79,Fur88} to the field
dependence of the PL line position the effective average concentration of Mn
in the digital alloy has been estimated to be $\bar{x}$ = 0.015.

The samples have also been characterized by electrical transport
measurements.
%\old{ *****SdH
At low temperatures pronounced Shubnikov-de-Haas
(SdH) oscillations and very well resolved quantum Hall plateaus
have been observed. 
%}
%\VB{to delete this phrase: it is incorrect to say about plateaus in R$_{xy}$ (which mean QHE) and SdH only (which are pre-QHE) in R$_{xx}$ }

 Two characteristic dependencies  measured in Hall bar and van-der-Pauw geometries in (Cd,Mn)Te structures
are shown in Figs.~\ref{transport}(a) and~\ref{transport}(b), respectively. 
While the amplitude of the oscillations increases in CdTe and (Cd,Mn)Te structures at rather high temperatures with 
raising magnetic field, the amplitude of the oscillations in (Cd,Mn)Te at low temperatures is a more complex function of the magnetic field $\bm B$ and shows a beating like pattern. 
The latter is clearly seen in Fig.~\ref{transport}(a) for $T= 1.6$ K ($B >3.5$~T) and in panel (b) for 4.2~K ($B
>4.5$~T). Such a behavior in diluted magnetic
semiconductor (DMS) structures  is well known to be caused by the exchange
interaction of electrons with Mn$^{2+}$  ions resulting in the
exchange-enhanced Zeeman
splitting~\cite{Kneip2006,Fur88,Dietl,DMS2010}.

\subsection{Dual grating top gate structure}
\label{imp}

For the ratchet effect experiments  we fabricated a dual grating top gate (DGG) superlattices on top 
of the QW structures.
A sketch of the gate fingers and a corresponding optical micrograph are shown in Figs.~\ref{structure}(b) and~\ref{structure}(c), respectively.

The grating-gate supercell consists of two metal stripes having
widths $d_1$ and  $d_2$  and spacings  $a_1$ and
$a_2$. The widths of the thin stripes were either $d_1 = 1.7$ or
1.85~$\mu$m, that of the wide stripes $d_2 = 3.7$~$\mu$m, see
Tab.~\ref{tab1}. Spacings in most of the structures were $a_1 =
2.8$~$\mu$m and $a_2=5.6$~$\mu$m.
To clarify the influence of the superlattice parameters on the ratchet effect,
we additionally fabricated one DGG structure with substantially different spacings $a_1 = 3.5$~$\mu$m and $a_2 = 7.0$~$\mu$m between the stripes.
The supercell is  repeated $N$ times to produce an asymmetric superlattice with the period $d = d_1 + a_1 +d_2 +a_2$, see Refs.~\cite{Olbrich_PRB_11,ratchet_graphene,staab2015}.
The two subgrating gates, each formed by  stripes of identical width,
can be biased independently. For this purpose all thin (top gate G1) and thick  (top gate G2) grating stripes have been connected by  additional gold stripes, see thick horizontal lines in
Fig.~\ref{structure}(c). The benefit of this geometry  is that  the periodic lateral electrostatic potential of the DGG structure
can be  varied in a controllable way.
Several $4\times 10$ and $4\times 4$\,mm$^2$ size  samples of the same batch were prepared.
The DGG structure has been fabricated on one half of the samples so that the other
half of the sample surface remains unpatterned  serving as reference part.
It has a total area of $w_2 \times N d$ where $w_2$ is the overlap
length 
%\fap{DW is has marked "`overlap length"' with "`?"' } 
of two gate lattices being of the order of hundreds of micrometer.

\begin{table}
 \begin{tabular}{|c|c|c|c|c|c|c|}
  \hline
  Sample                                & \#1 \#2               & \#3  & \#4  & \#5 &\#6 \\   % 2 = 5 , tauscht
  \hline
      \hline
  material                          & CdMnTe        & CdMnTe        & CdMnTe  & CdTe      &CdTe     \\
  \hline
  Au                                    & 25 nm                          & 30 nm & 15 nm     & 25 nm   &25 nm    \\
  \hline
  Dy                                        & -                          & -     & 75 nm     & -    &-     \\
  \hline
  $d_1$ ($\mu$m)                & 1.85                & 1.7         & 1.7      & 1.7  & 1.85      \\
  \hline
  $a_1$   ($\mu$m)          & 2.8                           & 3.5  & 2.8   & 2.8     &2.8     \\
  \hline
  $d_2$      ($\mu$m)           & 3.7                    & 3.7           & 3.7     & 3.7    &3.7    \\
  \hline
  $a_2$     ($\mu$m)        & 5.6                               & 7  & 5.8  & 5.8    &5.6     \\
  \hline
  $d$           ($\mu$m)        & 13.95                         & 15.9   & 14       & 14         &13.95    \\
  \hline
  $l$   ($\mu$m)                &  905                                & 880      & 875      & 875     &905     \\
    \hline
  $N$                                   & 65                                 & 56        & 63           & 63    &65    \\
    \hline
  $w_1$ ($\mu$m)                & 600                           &  450           & 450        & 450     &600    \\
    \hline
  $w_2$ ($\mu$m)                & 200                           & 350           & 350         & 350     &200     \\
    \hline
 \end{tabular}
 \caption{Sample parameters shown in detail in Figs.~\ref{structure}(b) and~\ref{structure}(c)}.
 \label{tab1}
\end{table}

In all samples, besides the sample \#4, the gate fingers have been made by electron beam lithography and subsequent deposition
of 25 or 30~nm thick gold films.
%Note that  these semitransparent in terahertz (THz) range.
The DGG structure of sample \#4 has been fabricated by depositing $75$ nm thick 
Dy and $15$ nm thick Au films.
%While not used in the present work, in general,
%\old{Making use p} 
Pre-magnetizating the hard magnetic Dy-based superlattice
%\old{this type of structures allows one to} 
enables realization of a magnetic ratchet 
%\old{effect caused by} \DW{
which features an inhomogeneous magnetic field, 
%\old{which was} 
recently suggested in~\cite{Budkin_Golub,JETP_Lett_review}.

\begin{figure}
    \includegraphics[width=\linewidth]{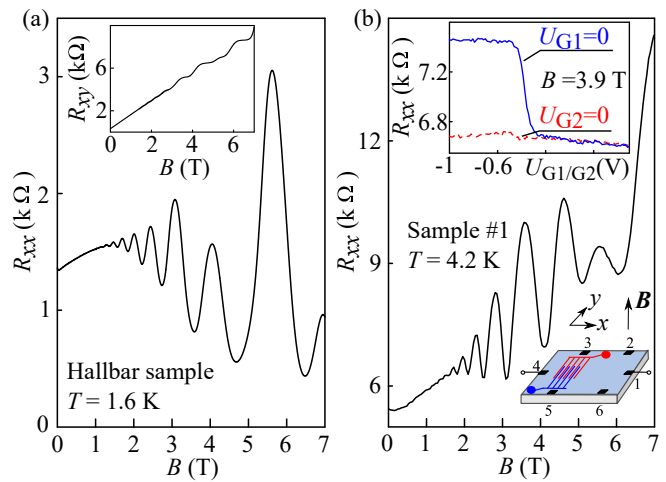}
    \caption{
		Magneto-transport experiments performed at
    temperatures $T= 1.6$ and 4.2~K in (Cd,Mn)Te structures.
(a) Unpatterned Hall bar sample. (b) Square shaped samples  with a DGG superlattice. 
For this measurement a van-der-Pauw geometry
has been used. Inset in (b) shows the dependence of the
longitudinal resistance, $R_{xx}$, on the gate voltage $U_{\rm
G1}$ ($U_{\rm G2}$) measured for $U_{\rm G2}=0$ ($U_{\rm G1}=0$).
%applied to the other gate.  
Measurements are presented for $B=3.9$ T corresponding to a maximum of the $R_{xx}$-oscillations.
Bottom inset in (b) sketches the DGG sample.
%\VB{SdH or QHE? i mean that at B$>3T$ one can see clear plateaus in R$_xy$ (inset in fig.2a) magnetoresistance? is Hall bar sampel shown?
}
\label{transport}
    \end{figure}

For the photocurrent and magneto-transport measurements  several  pairs of
ohmic contacts have been prepared, see inset in
Figs.\,\ref{transport}(b) and~\ref{gatedepB0}. 
%\sdg{A pair of point contacts deposited along $x$-direction
%in the middle of the edges allows us to study  photocurrent 
%in the direction normal to the grating fingers.
%Whereas two other pairs make it possible to measure the photocurrent 
%parallel to the grating fingers and to examine the unpatterned (reference) 
%part of the sample.}
%
%\fap{prefer the following text. pictures are changed. proglem with space for numbers}
%\old{Alternatively, but need extension of insets: 
Contact pads were placed in a way that the photo-induced currents 
can be measured perpendicularly  to the metal fingers ($J_x$, contacts 1 and 4) and parallel to them ($J_y$, contacts 3 and 5). 
Two additional contacts (2 and 6) were used for detecting the photocurrent signals from 
the unpatterned area  as a reference.  
Magneto-transport data for the (Cd,Mn)Te DGG structure at liquid helium
temperature are shown in Fig.~\ref{transport}(b). 
%\DW{comment: geometry of the sample, i.e. the position of the contacts between one measures is not clear!  - see printed text - }
%\sdg{Add here reference to Fig., geometry of the sample, i.e. position of contacts between one .. is not clear.
%see printed text??}
Note that the application of gate voltages to individual gates does not visibly influence  the period of 
$1/B$-oscillations of the longitudinal resistance, $R_{xx}$ (not shown).
%\VB{R$_xx$ has no period in B. So it would be better to say "`visibly impact on  B-dependence of R$_{xx}$"'}
%
This is due to the fact that the
area of DGG fingers is a very small fraction of the whole sample
area. Moreover, the variation of bias voltage applied to gate 1
($U_{\rm G1}$) also does not substantially affect the value of
$R_{xx}$ at a fixed magnetic field. The only detected deviation
from the value of $R_{xx}$ at zero gate voltages is an increase of
the resistance by $\sim 10$~\% for $U_{\rm G2} \approx -0.35$~V.
The gate voltage dependencies of the longitudinal resistance
measured at one of the maxima of the $R_{xx}$ oscillations 
are shown in the inset in Fig.~\ref{transport}(b).
%\VB{ it is not caused by SdH - see comment in the Capture
%magnetoresistance?}
%
The dependencies are obtained
at fixed magnetic field by sweeping the potential applied to
one gate and holding the other one at zero bias. The observed
behavior of $R_{xx}$ will be addressed below in the discussion of
photocurrent data. 
Note that  while the overall characteristic of the data obtained at different cooldowns is the same the onset of the resistance increase 
can be shifted on the gate voltage scale by $\pm 0.1$~V. This is ascribed to cooldown dependent charge trapping in the insulator.

%\FloatBarrier

\section{Experimental technique} \label{experi}

For THz excitation we applied a continuous wave (\textit{cw})
molecular optically pumped laser~\cite{hallgraphene,Kvon2008}. The
laser operated at the frequency $f = 2.54$ THz (photon energy
$\hbar\omega = 10.4$ meV,  wavelength $\lambda =118~\mu$m). The
incident power about 30~mW
%***power(910*200*10^-6*10^-6/((1.3*10^-3/2)^2*pi))=0.14, Sbeam = 1.3e-2 cm^2, s_dgg=1.82e-3 cm^2
was modulated at about $75$ or
625~Hz by an optical chopper. 
The radiation at normal incidence was
focused onto a spot of about $1.3$~mm diameter. The spatial beam
distribution had an almost Gaussian profile, measured by a
pyroelectric camera~\cite{Ziemann2000}. 
Taking
into account the size of the superlattice we obtain that the
%fraction of 
power irradiating the structure  is   
%***power
$P \approx 4$\,mW.
%(910*200*10^-6*10^-6/((1.3*10^-3/2)^2*pi))=0.14, Sbeam = 1.3e-2 cm^2, s_dgg=1.82e-3 cm^2
%
%\fap{$-1*(col(LOC.2X)/col(KI20101Data)/(100)/(2*50)/(2*\textbf{\textrm{0.0246}}/0.95))/((910*200*10^-6*10^-6/((1.5*10^-3/2)^2*pi)))$}
%
%***power
The radiation intensity and electric field strength on the sample are 
$I \approx 300$\,mW/cm$^2$ and $E_0 \approx 5.7$\,V/cm, respectively.
% 4 mW / 1.3e-2 cm^2 = 308 mW/cm^2, 2*Z = 2*120*3.14/3.5 = 215
%sqrt (0.3 W/cm2 * 2*Z V^2/cm^2) = sqrt (64.5V^2/cm^2) = 8 V/cm
Note  that almost all experimental data, except the scan
of the beam across the sample, are obtained for radiation focused
on the DGG superlattice.

The structures were placed in a temperature variable optical magneto-cryostat.
%with TPX \fap{DW marked TPX with "`?"'} windows.
The photocurrents were studied in the temperature range between 2 and 9~K.
The laser radiation was linearly polarized along the $x$-axis.
%\old{To explore the polarization dependence of
%the ratchet effect we 
%used crystal quartz  $\lambda/4$ plate and metal one-dimensional mesh polarizers.}
%***alpha
%To vary the azimuth angle $\alpha$ of linearly polarized radiation
%we obtained circularly polarized radiation by $\lambda/4$ plate
%and rotated a one-dimensional mesh polarizer placed behind the  plate.
%Note that for  $\alpha=0$ the radiation is linearly polarized along the $x$-axis.

The photoresponse is measured by the voltage drop $U$ across a load resistor $R_L = 50$~Ohm~$\ll R_s$ using standard lock-in technique. Here $R_s$ is the sample resistance.
%
%\old{The photoresponse is measured using standard lock-in technique in two
%electrical circuits by voltage 
%$U$ picked up across a load resistance $R_L\ll R_s$, where $R_s$ is the sample resistance.}
%\DW{comment on top of page 4. something like: "`Discussed is only one?"'}
The benefit of using of the small value of $R_L$ 
%$R_L\ll R_s$ 
is that the detected signal is unaffected
%not influenced 
by the sample resistance variation and is just proportional to 
the electric current generated by the THz radiation.
The current is calculated via $J = U/R_L$.

\section{Photocurrent at zero magnetic field}
\label{erstes}
%\FloatBarrier

\begin{figure}[h!]
    \includegraphics[width=\linewidth]{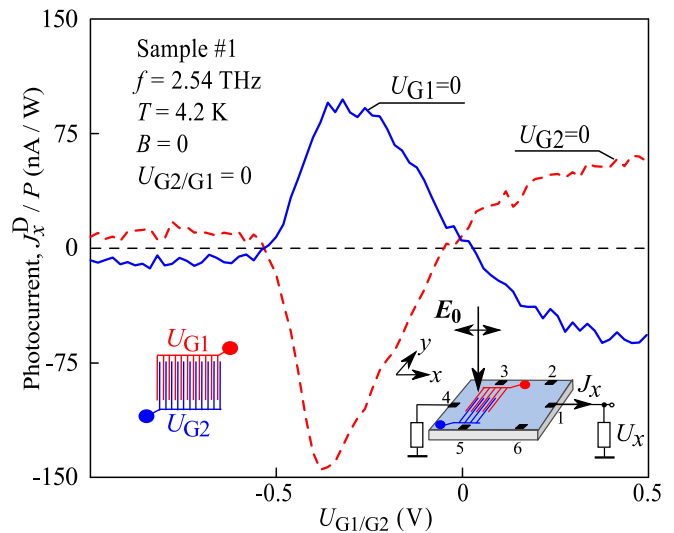}
\caption{
Dependencies of the normalized magnitude $J^{\rm D}_x/P$ on the gate voltage $U_{\rm G1}$ ($U_{\rm G2}$) applied to the stripes 
at zero potential of the other gate $U_{\rm G2} =0$ ($U_{\rm G1}=0$).
The polarization independent photocurrent contribution $J^{\rm D}_x$ is
extracted from 
the photocurrent polarization dependencies (not shown).
The data are obtained for (Cd,Mn)Te QW sample \#1
at zero magnetic field.
%and normalized to the radiation power $P$.
%These contributions to the total photocurrent $J_x$ are extracted from 
%the photocurrent polarization dependencies (not shown).
% ($\alpha = 0$).
%\VB{to change "`directed normal to the metal gate fingers $(\alpha = 0)$"' to "at  $\alpha = 0$"'   (same in the captures to figs. 6, 7, and 10)}
%Here and in further plots, 
Here and in the further plots, except for the one showing results on 
scan across the sample, the data are presented for the laser beam focused on the DGG superlattice.
Insets show  experimental geometries.
Here and in the further plots, the data are obtained for linearly polarized radiation with the electric field vector ${\bm E}_0 \parallel x$
directed perpendicularly to the metal gate fingers.
}
    \label{gatedepB0}
\end{figure}

We begin with briefly introducing the results obtained at zero magnetic field.
While the paper is devoted to the observation of
magneto-ratchet effect these results are important
for further analysis of the data on magnetic field induced photocurrents.

Illuminating the DGG superlattice at normal incidence we have observed
a photocurrent, whose magnitude and direction are sensitive to the
electrostatic potentials applied to the first (G1) and the second
(G2) gate sublattices. The photoresponse is detected
perpendicularly to the gate stripes as well as along them. The
fact that the photoresponce is generated by normally incident
radiation provides a first indication  that it is formed due to
the presence of the superlattice. Indeed it is well known that in
unpatterned (001)-oriented QW structures a photocurrent
signal is only detectable under oblique
incidence~\cite{DMSPRL09,DMS2PRB12}. The photocurrent consists of
a contribution independent of the radiation polarization
and a substantially smaller one 
dependent on the orientation of electric field
vector $\bm E_0$ of the linearly polarized radiation in respect to the
orientation of the gate stripes. This has been
demonstrated by measuring the photocurrent as a function the orientation of the $\bm E$-field vector in respect to the $x$-axis
%of the azimuth angle $\alpha$ 
(not shown). The same polarization behavior
has been previously observed for the ratchet effect in lateral GaAs-based structures~\cite{Olbrich_PRL_09,Olbrich_PRB_11} as well
as in DGG structures fabricated on top of InAlAs/InGaAs/InAlAs/InP
high electron mobility transistors (HEMT)~\cite{Otsuji_Ganichev}
and graphene~\cite{ratchet_graphene}. 

For unbiased gates, a non-zero
built-in electrostatic potential is formed due
to the presence of  metal stripes in the QW vicinity.
%\old{To be done: check built-in description.}
A direct evidence for the ratchet effect comes from the variation
of the photocurrent by application of the bias voltages to the
individual gates. 
%\fap{still no sentence:}
Indeed the ratchet effect is known to be
proportional to the averaged product 
\begin{equation} \label{Xi}
 \Xi = \overline{\frac{dV}{dx} |E(x)|^2} 
\end{equation}
of the derivative of the coordinate dependent electrostatic potential $V(x)$ and the distribution of the electric near-field  $E(x)$~\cite{Review_JETP_Lett,Nalitov,PopovAPL2013,PopovIvchenko}. 
In the following, we call the  parameter $\Xi$ the \emph{lateral asymmetry}.
%Consequently,  
The value of $\Xi$ may change sign depending on $V(x)$. Consequently
a variation of individual gate voltages should result in a change of 
the ratchet current including reversal of its direction, see the discussion in Sec.~\ref{discussion}.
%\old{May be already here, and not in two places marked later.
%To be done: remove words on built-in, but introduce potential inversion 
%with respect to the near field profile, reference to the discussion.}
%

Exactly  this behavior has been observed in the
experiment. 
%
%\DW{comment: Why is the asymmetry turned by setting the voltage of one gate to 0 and the other to finite value?}
%
In order to tune the lateral asymmetry, we applied
different  bias voltages $U_{\rm G1} \neq U_{\rm G2}$ to the
grating gates. Figure~\ref{gatedepB0} 
%\old{(a) and (b)} 
demonstrates the influence of the gate voltage variation on the amplitude and sign of 
%\old{the linear polarization dependent photocurrent ($J^A_x$) and}
%the one driven by the polarization independent radiation
the polarization independent photocurrent $J^D_x$~\cite{footnoteJD}.
%
%
%
%\old{, respectively.}   % hmmmmm
Holding one of the gates at zero bias and varying the gate voltage on 
the other one we could controllably change the lateral asymmetry.
%
%\old{To be done: remove words on built-in, but introduce potential inversion 
%with respect to the near field profile, reference to the discussion.}
%
The figure reveals that inversion of this lateral asymmetry obtained either by change of polarity of voltage applied to one gate or by interchange of gate voltage polarities applied to narrow and wide gates results in a change of sign of the photocurrent. 
This observation agrees well with the signature of ratchet currents: $J_x \propto \Xi$.
%characteristics for ratchet current, $J_x \propto dV(x)/dx$. 
%
Note that,  while the overall characteristics of the data obtained at different cooldowns is the same, the photocurrent sign inversions 
can be shifted on the gate voltage scale  by $\pm 0.1$~V. Similarly to the results addressed in Sec.~\ref{imp} this is ascribed to cooldown dependent charge trapping in the insulator. We also note that for large negative gate voltages all contributions vanish, which can be attributed to closing the transistor channel beneath the metal stripes of the DGG superlattice. 
%Note that for large negative gate voltages all contributions vanish, which can be attributed to closing the transistor channel beneath the metal stripes of the DGG superlattice.
Figure~\ref{gatedepB0} also reveals that at large positive gate voltages the photocurrent saturates  due to the effect of 
%the response of two-dimensional electron gas with 
high electron densities. 

To summarize, experiments at zero magnetic field provide a consistent picture demonstrating that the photocurrents are caused by the ratchet effect. They are (i)~generated due to the lateral asymmetry, (ii)~characterized by specific polarization dependencies for directions along and across the metal stripes, and (iii)~change the direction upon reversing the lateral asymmetry.
These results are in full agreement with the theory of ratchet effects excited by the polarized THz electric field in asymmetric lateral superlattices,  discussed below in Sec.~\ref{discussion}. The overall  behavior of the photocurrent is also in qualitative agreement with that of the electronic ratchet effects observed in semiconductor QW structures and graphene with a lateral superlattice~\cite{Olbrich_PRL_09,Olbrich_PRB_11,Review_JETP_Lett,ratchet_graphene}.

\section{Magnetic field induced photocurrent} \label{magnetic}
%\FloatBarrier

\subsection{Basic results}

\begin{figure}[ht]
        \includegraphics[width=\linewidth]{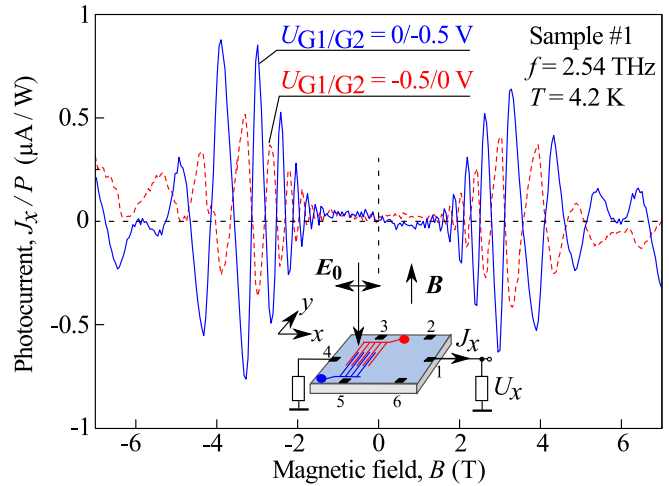}
        \caption{
Normalized photocurrent $J_x/P$
%, panel (a), and $J_y/P$, panel (b), 
as a function of the magnetic field $B$ measured in (Cd,Mn)Te
QW sample~\#1 for two combinations of the gate voltages. 
%The data are obtained for linearly polarized radiation and $R_L = 50$~Ohm. 
Solid blue line show the data
for the gate voltages  $U_{\rm G1} = 0$ and $U_{\rm G2} = -0.5$~V
and the dashed red line for $U_{\rm G1} = -0.5$~V and $U_{\rm G2}
= 0$, the setups denoted as $U_{\rm G1/G2} = 0/-0.5$~V and ${U_{\rm G1/G2} =
-0.5/0}$~V.
%
%\old{For briefness from here on we indicate the magnitudes of the
%gate voltages applied to lattices in the following way, e.g. for
%the former case  it will be labeled as $U_{\rm G1/G2} =
%0/-0.5$~V.}
%
The vertical dashed line indicates zero
magnetic field. The inset illustrates the experimental setup. }
  \label{jxjy4K}
    \end{figure}

\begin{figure}[ht]
        \includegraphics[width=\linewidth]{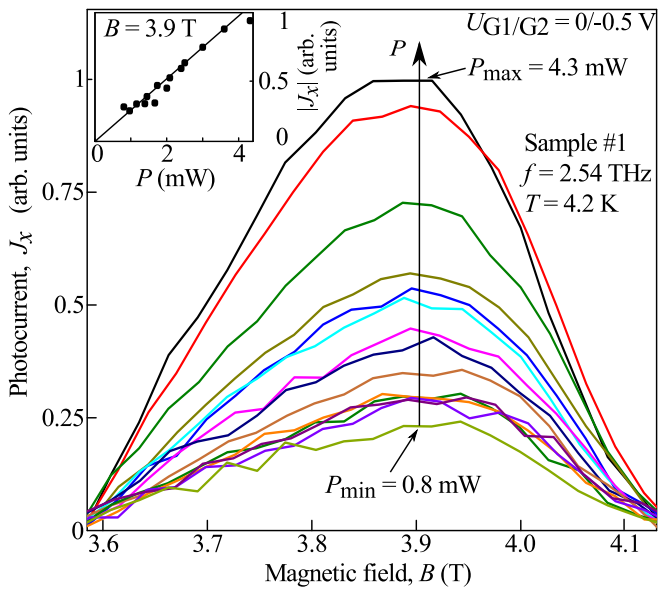}
        \caption{ Magnetic field dependencies of the photocurrent
        $|J_x|$ measured  at different levels of power ranging from 0.8 to 4.3~mW
        in (Cd,Mn)Te QW sample \#1. 
								%\old{The results 
								%are shown for the magnetic field range and 
								%demonstrate that the maxima position indicated by a vertical arrow ($B=3.9$~T)  does not change for various power levels.}
%				\VB{to delete "`are shown for the magnetic field range and"'}
								Inset shows  radiation power dependence of the photocurrent amplitude $|J_x|$ for $B=3.9$~T. Solid line is a linear fit.}
        \label{intensity}
    \end{figure}

Now we turn to the main part of the paper devoted to magnetic
field induced ratchet effects. Figure~\ref{jxjy4K} 
shows the normalized 
photocurrent 
%$J_x$ 
excited by linearly polarized
radiation 
%($\alpha = 0$) 
as a function of magnetic field 
applied perpendicularly
to the QW plane. The principal observation is that with
raising magnetic field the photocurrent drastically increases
%compared to the zero magnetic field ratchet photocurrent 
and, at
high magnetic fields, exhibits sign-alternating $1/B$-periodic
oscillations. Moreover, Fig.~\ref{jxjy4K} demonstrates that
the reversal of the lateral asymmetry results in
an inversion of the magneto-photocurrent direction providing a first clear
indication that it is driven by the ratchet effect. Measurements
of the magneto-photocurrent as a function of radiation power show
that it scales  linearly with the radiation power, i.e.
quadratically with the radiation electric field, $E_0$, see 
inset in Fig.~\ref{intensity}. 
Note that the positions of oscillation 
maxima/minima are independent of the radiation power, 
%we measured the magnetic field dependencies at different power levels, see 
Fig.~\ref{intensity}.
%and has substantially different magnitudes for positive and negative fields.
%{\color{red} However, for some measurements, e.g. probing the current in
%$y$ direction in sample~\#1, the detected photocurrent is even in
%$\bm B$ and has substantially different magnitudes
%for positive and negative fields, see Fig.~\ref{jxjy4K}(b). 
%%
%This, at a first glance surprising difference is in fact known for
%magnetic field induced photogalvanics excited in two-dimensional
%systems subjected to an external out of plane magnetic field.
%Opposite parities detected for magneto-photocurrents probed in a
%different in-plane direction are attributed to the Hall effect. 
%%
%It results in a deviation of the measured photocurrent
%direction from the original one due to 
%deflection by the Lorentz force~\cite{DantscherCR}. Thus in the
%following we will not pay much attention to the parity of the detected
%magneto-photocurrent.}
Similar results, particularly magneto-oscillations with current amplitudes
much larger than the photocurrent at $B=0$, are obtained for all studied samples.
Studying various samples at different experimental conditions we found that 
%in most cases the oscillating part of the photocurrent
%is odd in the external magnetic field and the magnitudes of
%oscillations detected for opposite magnetic field directions are
%close \spb{to each other}. Probing 
the current in the $y$-direction
%we observed that the detected photocurrent can also be even in $\bm B$ and 
is much smaller than $J_x$.
% (not shown).

\begin{figure}[ht]
    \includegraphics[width=\linewidth]{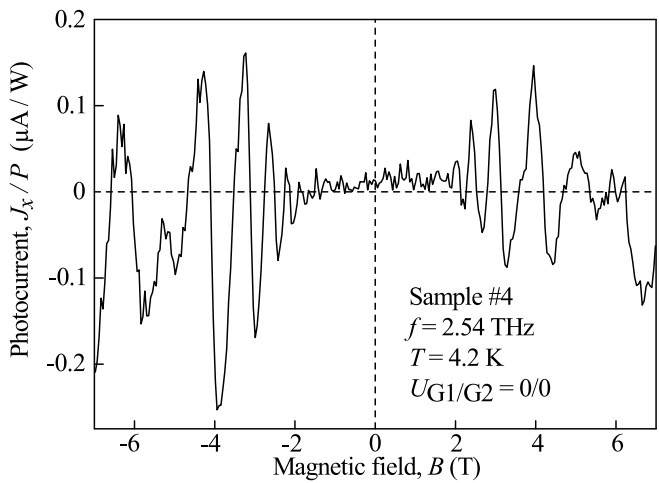}
    \caption{   Normalized photocurrent $J_x/P$ measured as a function of magnetic field $B$ in (Cd,Mn)Te QW sample~\#4 with the Dy grating.
%The data are obtained for linearly polarized radiation oriented
%normal to the metal gate fingers ($\alpha = 0$).
        }
    \label{DY}
\end{figure}

Data from the (Cd,Mn)Te QW sample~\#4 with ferromagnetic Dy gate
are shown in Fig.~\ref{DY}.
The period and magnitude
%\VB{to delete "`period and"'}
of the $1/B$-oscillations correlate with transport behavior and 
are very close to those shown for (Cd,Mn)Te QW samples with gold DGG structure presented for sample~\#1, Fig.~\ref{jxjy4K}.

\begin{figure}[h]
    \includegraphics[width=\linewidth]{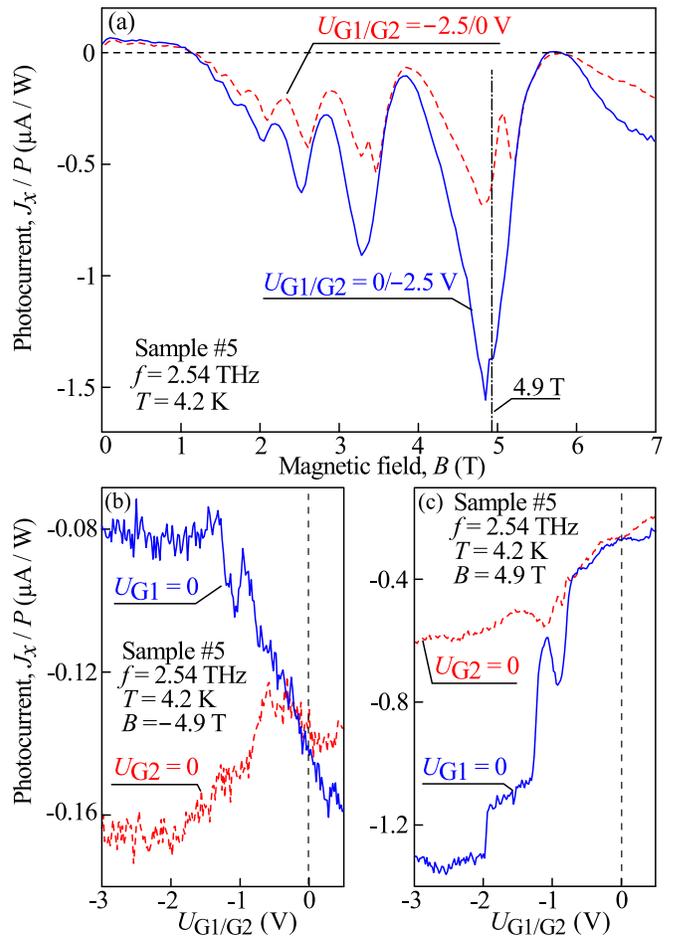}
    \caption{
        (a) Normalized photocurrent $J_{x}/P$ as a function of  magnetic field $B$  in CdTe-based QW sample~\#5  
				at two gate voltage sequences:  $U_{\rm G1}/ U_{\rm G2}= -2.5$~V$/0$ and $U_{\rm G1}/ U_{\rm G2}=0/$$-$2.5~V.
%The vertical dashed line represents zero magnetic field \textit{B}.
%The data are obtained for linearly polarized radiation directed normal to the metal gate fingers ($\alpha = 0$). 
Panels (b) and (c) show dependencies of the
photocurrent magnitude on the gate voltage $U_{\rm G1}$ ($U_{\rm G2}$) obtained for zero potential on the other gate $U_{\rm G2}$ ($U_{\rm G1}$). The data in panels (b) and (c)
are obtained for the photocurrent maxima/minima at $B=-4.9$
 and 4.9~T, respectively.
            }
    \label{CdTe}
\end{figure}

%
%\DW{comment: can you correlate the defect with different transport behaviour?}
%
Figure~\ref{CdTe} shows another example of magneto-photocurrent measured for nonmagnetic CdTe QW sample~\#5 with gold DGG.
While oscillations are clearly detected the signal is superimposed with a substantial background current.
Because of the background the photocurrent, in contrast to samples \#1 and \#4, doesn't change sign with varying magnetic field. The background photocurrent has also been  obtained for  (Cd,Mn)Te QW samples  \#2 and \#3. 
We attribute an appearance of the background  to imperfections of our large-size superlattices
which are present in some structures as confirmed by optical microscope images~\cite{footnoteimperfections}. 
The imperfections locally reduce symmetry and give rise to ratchet unrelated magneto-gyrotropic photogalvanic 
currents in the ``bulk'' of the QW, see e.g. Ref.~\cite{Belkov2008}. 
% In the following we 
%will focus in experimental part on the results obtained on (Cd,Mn)Te QW samples \#1 in which imperfections are almost absent.
%In the following we will focus on the results obtained on (Cd,Mn)Te QW samples \#1and \#4, in which imperfections are almost absent.}
%In the following we will focus on the 
%results obtained on (Cd,Mn)Te QW samples \#1and \#4, in which imperfections are almost absent.
%\DW{Question mark for the next sentence}
%Discussing the effect of lateral potential asymmetry, however, we return to CdTe QW sample~\#5 
%and (Cd,Mn)Te QW DGG samples with other geometrical parameters of DGG.

\subsection{Laser beam scan across the DGG structure}

\begin{figure}[ht]
        \includegraphics[width=\linewidth]{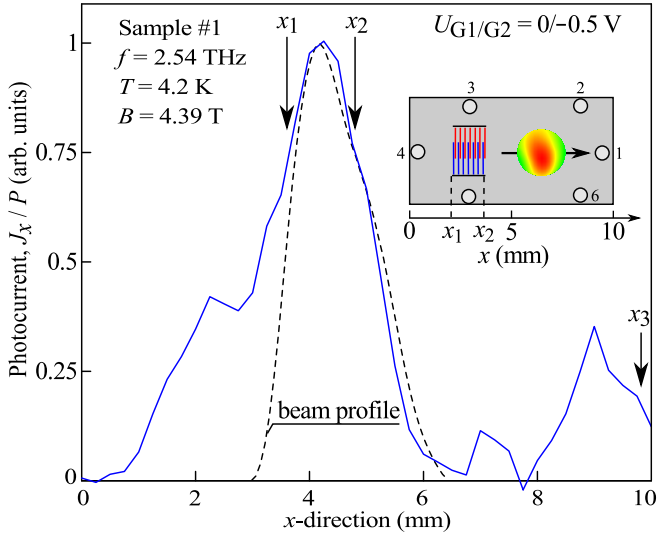}
        \caption{
            Laser spot position dependence of the normalized photocurrent $J_x/P$ measured in (Cd,Mn)Te QW sample~\#1  at a maximum of the photocurrent oscillation at  $B=4.39$~T and   $U_{\rm G1} = 0$, $U_{\rm G2} = -0.5$ V.
Inset shows the setup with 
%\old{
the laser spot scanned along the $x$-axis.
% and the current is picked up from source-drain contact pair.}
%\VB{and $J_x$ is picked up (term "`source-drain seems incorrect here - indeed, what does it mean "`source"'?)}
Positions $x_1$ and $x_2$ correspond to the borders of the superlattice whereas $x_3$
denotes the sample edge position. The dashed line represents the laser beam spatial distribution which is measured by a pyroelectric camera  and scaled to the current maximum.
        }
        \label{Scan}
    \end{figure}

To prove that the magneto-ratchet effect stems from the irradiation of the superlattice we scanned the laser spot across the sample along the $x$-direction. The photocurrent $J_x$ was measured for  $B=4.39$ T and the gate voltage combination $U_{\rm G1/G2}=0/-$0.5~V, i.e. at a maximum of one of the magneto-oscillations, Fig.~\ref{jxjy4K}. The  photocurrent generated by linearly polarized radiation with $\bm E_0 \parallel x$ as a function of the radiation spot position and the corresponding experimental geometry are shown in Fig.~\ref{Scan}. The photocurrent reaches its maximum for the laser spot centered at the superlattice and rapidly decays with the spot shifting away. For the beam spot touching the sample edge 
%close to the sample edges 
the current increases again. This is caused by an edge photocurrent~\cite{footnotescan} known for graphene~\cite{Karch_PRL_2011,Glazov2014} and semiconductor QWs~\cite{NTTI2014}. 
%\fap{sentence quite similiar in text and in references}
%\VB{do we need ref. [53] (our condmat)? }
Comparison with the laser beam spatial distribution measured by a pyroelectric camera (dashed curve in Fig.~\ref{Scan}) shows that the change of the photocurrent  for the scans across the DGG structure only slightly deviates from the Gaussian intensity profile. Note that the DGG area is smaller than the beam spot.
These findings unambiguously demonstrate that the photocurrent is caused by irradiation of the superlattice.

\subsection{Effect of the lateral asymmetry variation of the photocurrent}
\label{osci}
%\FloatBarrier

\begin{figure}[ht]
        \includegraphics[width=\linewidth]{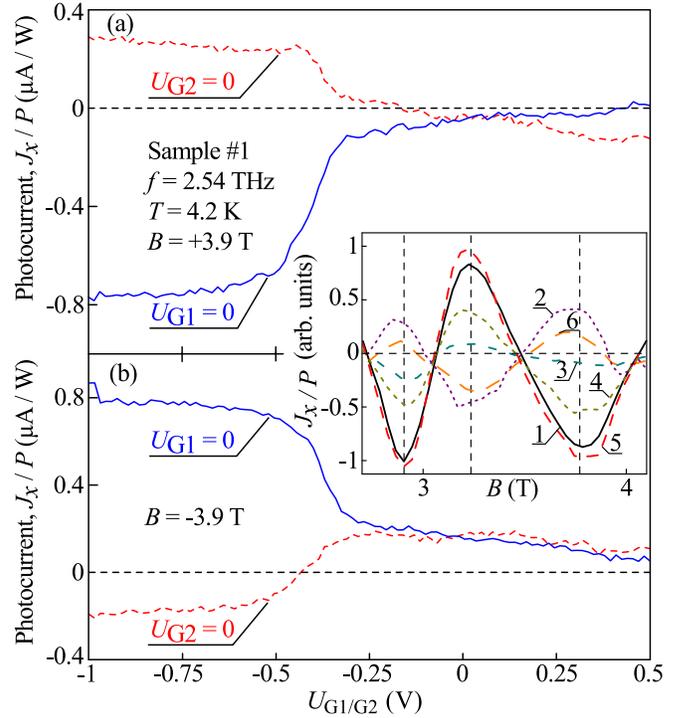}
\caption{Dependencies of the photocurrent magnitude on 
$U_{\rm G1}$ ($U_{\rm G2}$) obtained for zero potential on
the other gate $U_{\rm G2}$ ($U_{\rm G1}$). The data are obtained
for 
%the photocurrent measured on 
(Cd,Mn)Te QW sample~\#1  
at $B=3.9$~T  and $-3.9$~T, panels (a) and (b), respectively. The inset shows the magnetic field
dependence of the photocurrent measured for different gate voltage combinations. 
Vertical dashed lines demonstrate that the maxima/minima positions
%\DW{DW: maximum positions - marked by DW} 
of the magneto-photocurrent are not shifted for all gate voltage combinations. 
The numbers in the inset correspond to:
1 $\rightarrow$ $U_{\rm G1/G2}=0/-0.5$~V; 2 $\rightarrow$  $U_{\rm
G1/G2}=-0.5/0.5$~V; 3 $\rightarrow$  $U_{\rm G1/G2} = 0.5/0$~V; 4
$\rightarrow$  $U_{\rm G1/G2} = -0.5/-0.5$~V; 5 $\rightarrow$
$U_{\rm G1/G2} = 0.5/-0.5$~V and 6 $\rightarrow$  $U_{\rm G1/G2} =
-0.5/0$~V.                   }
        \label{asymmetry}
    \end{figure}

In order to explore the role of the lateral asymmetry, a prerequisite for the ratchet photocurrents, we have systematically studied
magneto-photocurrents for different combinations of the gate voltages. The
results are shown for different samples in Figs.~\ref{gatedepB0},~\ref{jxjy4K},~\ref{CdTe} and~\ref{asymmetry}.

While the magnitude and the sign of the
oscillations are strongly affected by the variation of the gate
potentials the maxima/minima  positions remain unchanged. 
%This has been
%already shown for selected combinations of gate  voltages in
%Figs.~\ref{jxjy4K} and~\ref{CdTe} and confirmed for DGG structure
%\#3 fabricated with 30\% larger spacings between metal fingers
%($a_1$ and $a_2$), where the only difference observed was a
%reduction of the signal magnitude (not shown). 
To show that this
conclusion is valid for the whole range of gate potentials used in
our measurements we studied the photocurrent for a large set of
different magnitudes and polarities of $U_{\rm G1}$ and $U_{\rm
G2}$. These results are presented in the inset in
Fig.~\ref{asymmetry} for $B$ ranging from 2.7 to
4.1~T.

Using the fact that maxima positions do not shift upon variation
of $B$ we fixed the magnetic field and controllably varied the
lateral asymmetry by changing the potential applied to
one of the gates and holding the other one at zero bias. 
Figures~\ref{asymmetry}(a) and~\ref{asymmetry}(b) show the results obtained for (Cd,Mn)Te
QW sample~\#1 at magnetic field strengths 3.9 and $-3.9$~T,
respectively. The figures reveal that the polarity of photocurrent
changes with the sign of the asymmetry parameter $\Xi$, Eq.~\eqref{Xi}.
%polarity of $dV(x)/dx$. 
This is
primarily seen by comparing the curves obtained by
variation of the potentials applied either to the thin (dashed lines) or
thick (solid lines) gate stripes. 
The observed stronger change of
the signal for $U_{\rm G1} (U_{\rm G2}) \approx -0.35$~V  
corresponds to an increase of the longitudinal resistance
detected in magneto-transport measurements, 
see inset in Fig.~\ref{transport}(b).
%\VB{The strong change of the photocurrent at variation of $U_{G2}$ potential corresponds to  gate-voltage dependence of the longitudinal resistance, see the ibset in Fig.2 (b) and ...}
%
%\pop{I do not understand this sentence at all. Starting from HERE: }
It should be noted 
%Note also 
that, due to the abovementioned built-in potential,
the ratchet current can be detected even for zero gate voltages at
both gates.
%due to the built-in potential caused by the presence
%of  metal stripes, reversing sign of the applied voltage does not
%lead automatically to sign inversion  of the electrostatic
%potential acting on the electrons. 
%\pop{:END}
%
%\old{To be done: remove words on built-in, but introduce potential inversion 
%with respect to the near field profile, reference to the discussion.}

The same behavior upon variation of the lateral asymmetry has been 
observed for Dy-based DGG sample~\#4.
Studying the photocurrent as a function of the gate voltage
in CdTe QW sample~\#5 and in (Cd,Mn)Te samples~\#2 and~\#3
we obtained, that, despite the background, 
the photocurrent magnitude is also controlled by the lateral asymmetry.
Corresponding results obtained for CdTe QW DGG structure~\#5
at $B = \pm$4.9~T are shown in Figs.~\ref{CdTe}(b) and~(c).

\begin{figure}[ht]
    \includegraphics[width=\linewidth]{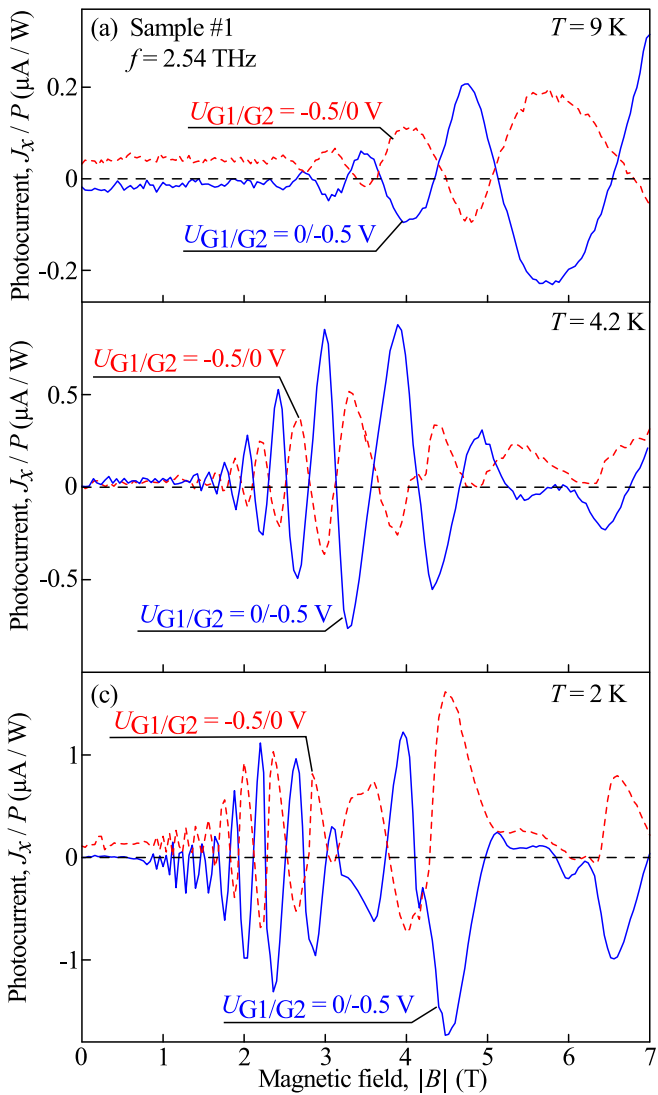}
    \caption{   Normalized photocurrent $J_{x}/P$  as a function of 
		$B$ measured in (Cd,Mn)Te QW DGG sample~\#1 for three temperatures and
    two gate voltage combinations. 
		%The data are obtained for linearly polarized    radiation directed normal to the metal gate fingers ($\alpha = 0$).     
		}
    \label{temperatur}
\end{figure}

Now we turn to the analysis of the $1/B$-oscillations of the
photocurrent which are obviously related to the
SdH oscillations.
%*****SdH
%\VB{are we sure that at high B these oscillat. are indeed SdH?(what about QHE?) 
%should we use term "magneto-resistance oscillations"?)
%This is a question also for some later statement using term "SdH"}
%\fap{Die Frage ist ja schon geklärt, wollte nur nochmal darauf hinweisen, auch "`period"' und "`SDH"' weiter unten hat VB noch angemerkt.  }
Comparison with the magneto-transport data reveals that the period of the $1/B$-oscillations is
indeed equal to that of the longitudinal magneto-resistance oscillations. 
This result  is confirmed for all samples and temperatures used in our experiments.
In the CdTe samples and (Cd,Mn)Te sample at relatively high temperatures 
the amplitude of oscillations grows with the magnetic field, see e.g. Fig.~\ref{temperatur}. 
% This is shown in Fig.~\ref{temperatur}(a) for  (Cd,Mn)Te sample \#1 at $T = 9$~K. 
By contrast, 
%This behavior, however, strongly differs 
in diluted magnetic semiconductor (Cd,Mn)Te QW DGG samples
%the behavior at lower temperatures, 
%For (Cd,Mn)Te QW DGG samples 
%where 
%we observed that 
at liquid helium temperature $T=4.2~K$ the rise of the magneto-photocurrent amplitude at low fields is followed by its
reduction at higher fields, see Fig.~\ref{temperatur}(b).
At the lower temperature $T=2$~K this reduction, now detected at even lower magnetic fields, 
is followed by a further increase of the photocurrent magnitude at higher $B$, see data in
Fig.~\ref{temperatur}(c).
Similar ``beats'' of the magneto-oscillation amplitudes are
detected for the longitudinal resistance, 
Fig.~\ref{transport}, in agreement with the presence 
of the exchange enhanced Zeeman splitting in DMS materials~\cite{Fur88}.  The
enhanced Zeeman splitting is caused by the exchange interaction
between electrons and Mn$^{2+}$  ions 
%in diluted magnetic semiconductors and is 
and described by the modified Brillouin
function. The spin splitting first grows linearly with $B$ 
and then saturates. The temperature increase leads to a decrease of exchange interaction and
shifts  the  saturation to higher fields. All these features are
detected in both transport and magneto-photocurrent experiments.
Moreover, due to the enhanced spin splitting the  oscillations measured
at low temperatures and high magnetic fields 
become spin-split which results in the beats.
%All the above observations demonstrate that the observed
%ratchet oscillations are caused by
%the oscillating  longitudinal resistance, \sdg{which,
%% caused by SdH effect, 
%%\DW{i.e. the SdH effect,}
%for (Cd,Mn)Te samples at low temperatures and high
%magnetic fields, are affected by substantial} contributions of the 
%exchange enhanced Zeeman splitting.

%\FloatBarrier
\section{Discussion}
\label{discussion}

%\fap{VB: is it correct statement? i mean that same one is valid for structures without lateral superlattice }

In low-dimensional semiconductor structures with lateral superlattices,
a $dc$ electric current is generated due to the action 
of an electromagnetic wave's $ac$ electric field~\cite{JETP_Lett_review}.
The effect of the superlattice is twofold:
It generates a one-dimensional periodic electrostatic potential $V(x)$ acting upon the 2D carriers with $x$ being the
superlattice principal axis, and causes a periodic spatial modulation of the THz electric field due to the near field diffraction.

\begin{figure}
\includegraphics[width=\linewidth]{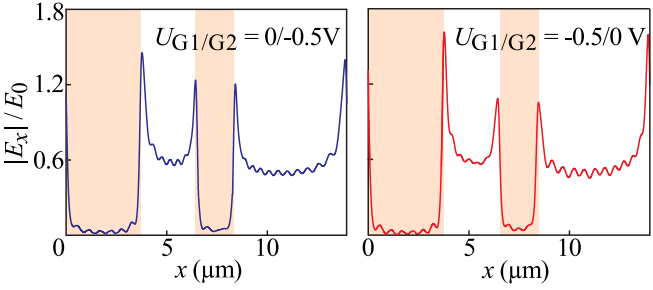}
\caption{Electric near-field $E_x(x)$ normalized to the electric field
of the incident THz wave. Shaded regions correspond to the
positions of DGG metal fingers. 
%\old{To be done: give parameters used for the calculations.}
Results are obtained for parameters of (Cd,Mn)Te QW sample \#1 with dielectric constant $\epsilon = 10$ at the
frequency 2.54~THz. 
The momentum relaxation time $\tau$ is obtained from the  
electron mobility specified in Sec.~\ref{sample}.
} \label{field}
\end{figure}

Figure~\ref{field} shows coordinate dependence of the THz electric near-field $E_x(x)$ calculated
for the  (Cd,Mn)Te QWs based DGG structure
for two combinations of the gate voltages.
The electric field distribution caused by the near-field diffraction
is calculated  for  radiation with frequency $f=2.54$~THz applying
a self-consistent electromagnetic approach based on the integral equation
method described in detail in Ref.~\cite{Fateev2010}.
%The curves are obtained for 
%the structure period 
%for DGG parameters  relevant to samples~\#1, \#2, \#4 -- \#6.

\begin{figure}
\includegraphics[width=\linewidth]{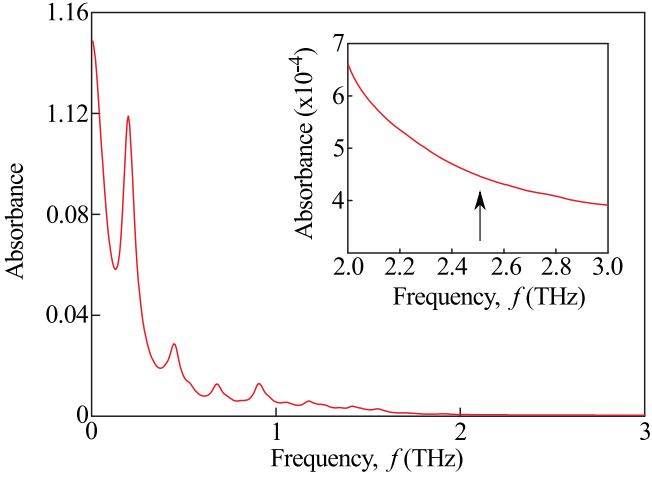}
\caption{Terahertz absorption spectrum of the structure. The inset
shows the spectrum in the vicinity of the frequency 2.54~THz (indicated by vertical arrow) used in our
experiments. 
%\old{To be done: give parameters used for the calculations.}
%The parameters used in the calculations are the same as for Fig.~\ref{field}.
The curve is calculated for the parameters used in Fig.~\ref{field}.
} \label{Absorption}
\end{figure}

Figure~\ref{Absorption} shows the calculated THz absorption
spectrum of the structure. It is seen that the plasmonic
resonances in the QW are excited at frequencies well below the
operation frequency of 2.54~THz. Therefore in the experimental frequency range the absorption follows the
Drude law, 
see the
inset in Fig.~\ref{Absorption}. As a result, the near-field
distribution in the QW shown in Fig.~\ref{field} is caused almost solely by
the metal dual-grating gate and not by the plasma oscillations.
%in the QW and 
Hence one can neglect the dependence of the near field
%hence it does not depend 
on the voltage
applied to the DGG. Moreover, since the radiation frequency is
much higher than that of plasmonic resonances we consider only
electronic ratchet mechanism below.

%replaced************************
\begin{figure}
        \includegraphics[width=\linewidth]{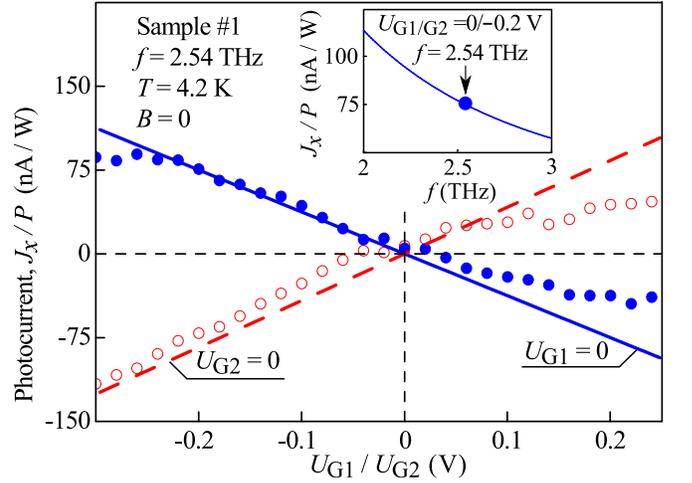}
        \caption{Calculated  dependencies of the photocurrent 
				$J_x = j_0 w_2$ on the gate voltages $U_{\rm G1}$ (dashed line) and $U_{\rm G2}$ (solid)
				fitted to experimental curves shown in Fig.~\ref{gatedepB0} ($w_2$ is defined in Fig.~\ref{structure}).
				The lines are obtained after Eq.~(\ref{j_zero_B}).
%				using parameters relevant to the corresponding experiment 
%				on (Cd,Mn)Te QW sample~\#1, see Fig.~\ref{gatedepB0}.
				%\old{To be specific we used, in accordance with experimental conditions, we used 
				%for calculations ${\bm E}|| x$.}
				To be specific, we used for calculations ${\bm E}_0 \parallel x$ in accordance with the experimental conditions. 
	      For the calculations we used the (Cd,Mn)Te DGG QW structure parameters discussed in 
				Sec.~\ref{sample}, electric near-field $E_x(x)$ shown in Fig.~\ref{field},  
				lattice temperature $T = 4.2$~K, 
				%electron density $n_e= 4.5 \times 10^{11}$~cm$^{-2}$ at $U_i=-0.2~V$, 
				frequency $f = 2.54$~THz and effective mass $m = 0.1 \, m_0$.
				%, see Ref.~[\onlinecite{Dremin2005}].
				The only %$y-$scaling 
				fitting parameter used here is the relaxation time 
				$\tau_\varepsilon = 4$~ns.  
				Inset shows the calculated frequency dependence of the photocurrent $J_x$ 
				together with the experimental result for $f = 2.54$~THz (dot).
				%calculated
				%for one combination of gate voltages. Dot shows experimental result for the
				%same gate voltage combination.
			}
				        \label{PopovFateev}
    \end{figure}

The ratchet current is generated by a combined action on carriers
of both the radiation near field, transmitted through the grating,
and the periodic static potential
$V(x)$~\cite{Olbrich_PRL_09,Review_JETP_Lett,Nalitov,Olbrich_PRB_11,ratchet_graphene}.
%
%\DW{comment: discussion .... offset .... at VG=0 ?? to the whole paragraph:}
%
A crucial condition for $dc$ electric current generation is
that the lateral superlattice is asymmetric.
%$V(x)$ and the near field intensity $|E(x)|^2$ are to be \sdg{asymmetric. \textbf{(more precise formulation is needed)}}
% and phase shifted to each other.} \fap{ELLGGB: "`asymmetric and"' remove}
%\fap{is that the lateral superlattice is to be asymmetric.}
%Flipping the lateral asymmetry applying 
%voltages of different magnitudes and signs to the subgratings 
%reverses the periodic potential $V(x)$
%in the QW with respect to the near-field distribution.  
Applying the voltages $U_\text{G1}$ and $U_\text{G2}$ in a controllable way, one can vary the potential $V(x)$ and reverse the lateral asymmetry $\Xi$.
Consequently, the ratchet photocurrent can also reverse its sign as observed in experiment, see Fig.~\ref{gatedepB0}. 
%
%\old{To be done: may be here to say words on 
%symmetry of electrostatic and near field potentials. 
%introduce potential inversion 
%with respect to the near field profile, reference to the discussion.}

The ratchet
current has components both perpendicular and parallel to the metal fingers and consists, in general, of  
polarization-independent, linear- and circular-ratchet
contributions. 
All these individual ratchet currents have been
observed and described in various semiconductor-based
structures~\cite{Olbrich_PRL_09,Review_JETP_Lett,Olbrich_PRB_11}
as well as in graphene with  lateral periodical top
gates~\cite{Nalitov,ratchet_graphene}.
The mechanism leading to photocurrent formation can be
illustrated on the polarization-independent
photocurrent caused by the Seebeck ratchet (thermoratchet) effect. This
contribution dominates at zero magnetic field (Sec.~\ref{erstes}, Fig.~\ref{gatedepB0}) and, as we show below, is
most relevant to the observed magnetic ratchet current. The Seebeck
ratchet effect is caused by inhomogeneous heating of two-dimensional electron gas and subsequent 
relaxation of the electron temperature. Due to the
near-field space modulation, 
%the magnitude of 
the field $E(x)$ acting upon electrons 
%is modulated in space, i.e. the radiation
%field $E(x)$ 
depends on the coordinate $x$, Fig.~\ref{field}.
The field heats the electron gas changing the effective electron
temperature locally to $T(x) = T + \delta
T(x)$. Here we assume that, within a short time, the electron distribution becomes a quasi equilibrium Fermi-Dirac  distribution with the temperature $T(x)$. 
The space-modulated temperature correction is defined by the energy
balance and can be found from the following equation~\cite{Nalitov}
\begin{equation}
    {k_\text{B}\delta T(x) \over \tau_\varepsilon} = 2 |E(x)|^2 {e^2\over m\omega^2\tau}.
\end{equation}
Here $k_\text{B}$ is the Boltzmann constant, $\tau_\varepsilon$ is the 
temperature relaxation time, $\tau$ is the momentum relaxation time, $e$ is the electron charge, and  $m$
is the electron effective mass.
We take into account that, in the present experiment
${\omega\tau \gg \omega_c\tau \gg 1}$, where $\omega_c=|eB|/(mc)$
is the cyclotron frequency.

As a result of the inhomogeneous heating, electrons diffuse from warmer to 
colder regions, and form a nonequilibrium density profile given by
%
%\old{Check for $-{\pi^2\over 6}$ or over 3
\begin{equation}
\label{delta_N}
	{\delta N(x) \over N} = -{\pi^2\over 6} \left( {k_\text{B}T \over \varepsilon_\text{F}} \right)^2 {\delta T \over T},
\end{equation}
%}
%
where $N$ is the average electron density, $\varepsilon_\text{F}$ is the Fermi energy.
The ratchet current at zero magnetic field can be represented as a drift current of the
electrons in the electric field $\mathcal{E}$ of the space
modulated electrostatic potential $V(x)$. From Ohm's 
law $j = \sigma \mathcal{E}$ with $\mathcal{E} = -(1/e)dV/dx$ we have 
%
%\old{Check for $-{e \tau \over m}$ or with $e^2$
\begin{equation}
	j_0 = -{e \tau \over m} \overline{\delta N(x) {dV \over dx}},
\end{equation}
%}
%
where the bar denotes averaging over the superlattice period.
Finally, the ratchet current density at  zero magnetic field is 
given by~\cite{Nalitov,JETP_Lett_review}
%
%\old{Check for factor 2 or not 
\begin{equation}
\label{j_zero_B}
	j_0 = {\pi e^3 \tau_\varepsilon k_\text{B}T \over 3m\hbar^2\omega^2\varepsilon_\text{F}} \Xi,
%	\overline{ \frac{dV(x)}{dx} |E(x)|^2 }.
\end{equation}
%}
where $\Xi$ is given by Eq.~\eqref{Xi}.
 
Figure~\ref{PopovFateev} shows the  photocurrent calculated after Eq.~\eqref{j_zero_B}
together with the data of Fig.~\ref{gatedepB0}. The result of Fig.~\ref{PopovFateev} 
is obtained for the near-field $E_x(x)$ shown in Fig.~\ref{field}.
Parameters used for the calculations correspond to the sample~\#1.
%\old{Ivchenko: Do not understand the meaning of this sentence: 
%are relevant to the corresponding experiment 
%on (Cd,Mn)Te QW sample~\#1 being obtained from independent measurements. }
%We compare the experimental data on zero-field ratchet current with calculations. The results presented in Fig.~{[Fateev]} show a good agreement in both ...dependence and absolute values. 
The only adjustable parameter is the temperature relaxation time taken as $\tau_\varepsilon = 4$~ns.
%(at $T=4.2$~K, $\varepsilon_\text{F}=8.8$~meV for the concentration $n_e=4.5 \times 10^{11}$~cm$^{-2}$, and  $\varepsilon_\text{F}=12.9$~meV for $n_e=6.6 \times 10^{11}$~cm$^{-2}$). 
This value agrees with 
%the experimental and calculated 
the temperature relaxation time of few nanoseconds
reported  for a degenerate electron gas in GaAs-based heterojunctions~\cite{DasSarma1990}. 
%Some difference can be attributed to a difference in material and in a stronger confinement in our quantum wells in comparison with heterojunctions.
Figure~\ref{PopovFateev} shows qualitative and reasonably good  quantitative agreement between theory and 
experiment. 
%\old{To be done: may be here to say words on 
%symmetry of electrostatic and near field potentials. 
%introduce potential inversion 
%with respect to the near field profile, reference to the discussion.}
%\old{It is worth noting that non-zero ratchet current appears only if 
%the profile of the electrostatic potential in QW is asymmetric in respect 
%to the profile of the electric near-field in such a way that there is 
%no mirror symmetry in the product $(dV(x)/dx)|E(x)|^2$. 
%Furthermore, the sign of the photocurrent changes for opposite 
%asymmetries of the electrostatic potential in respect to 
%the profile of the electric near-field.  
%In described experiments this the lateral asymmetry is controlled 
%by reversing gate voltages applied to gates~1 and~2, see Fig.~\ref{PopovFateev}.
%}
%It is worth noting that non-zero ratchet current is proportional to 
%the parameter $\Xi$ of lateral asymmetry of the DGG defined by Eq.~(\ref{asymmetry}).
% $\overline{{dV / dx} |E_0(x)|^2}$ describing the lateral asymmetry of the DGG. 
%In the reported experiments, the the value of $\Xi$ %lateral asymmetry 
%is controlled applying voltages to gates~1 and~2. In particular, the 
%photocurrent direction is reversed when changing the 
%$dV/ dx$ polarity, see Fig.~\ref{PopovFateev}.

The above analysis suggests that, in the absence of magnetic field, the ratchet current is rather weak for
degenerate electrons because, according to Eq.~\eqref{delta_N},
%However, the degenerate electrons are heated not very effectively: 
the relative concentration correction $\delta N/N$ is much smaller than the temperature correction $\delta T/T$.
%by a factor $(\pi^2/3)(k_\text{B}T/\varepsilon_\text{F})^2 \ll 1$. 
%
%
Below we show that the situation changes drastically in the presence of a quantizing magnetic field $\bm B \parallel z$.

The magnetic field induced ratchet currents 
in the directions perpendicular, $j_x$, and parallel, $j_y$, to the stripes
of the superlattice are given by~\cite{JETP_Lett_review}
\begin{equation}
    \label{j}
    {j_x\over j_0} = {48\varepsilon_{\rm F}^2\over \hbar\omega_c k_\text{B}T }
		{(\omega_c\tau)^2\over [1+(\omega_c\tau)^2]^2}
		{\sinh{z}-z\cosh{z} \over \sinh^2{z}} \delta,
\end{equation}
\begin{equation}
	    \label{jy}
	 j_{y}= -\frac{B_z}{|B_z|} {1+3(\omega_c\tau)^2\over 2(\omega_c\tau)^3} j_x.	
\end{equation}
Here the magneto-oscillations are described by the factor $\delta$ ($|\delta| \ll 1$)
\begin{equation}
\label{delta}
    \delta =
    \cos{ \left( {2\pi\varepsilon_{\rm F} \over \hbar\omega_c}  \right)}
    \cos{ \left(    {\pi\Delta_Z \over \hbar\omega_c} \right)}
    \exp{\left(-{\pi \over \omega_c\tau}\right)},  
\end{equation}
%
%\sdg{
%where the dependence on ${z= 2\pi^2 k_\text{B}T / (\hbar\omega_c)}$ describes the temperature
%damping of the magneto-oscillations. 
%The theory is developed in first order in $\delta$ considering the
%oscillations with the period determined by the Fermi energy 
%and generalize the theory developed in Ref.~\cite{JETP_Lett_review} to the case of two spin subbands 
%splitted by the energy $\Delta_Z$ due to Zeeman effect. }
%\old{Ivchenko: 
while the dependence on $z = 2 \pi^2 k_{\rm B}T/(\hbar \omega_c)$ describes the oscillation suppression with increasing temperature, Eqs.~\eqref{j_zero_B}-\eqref{jy} are derived in the first order in $\delta$.
As compared with the theory developed in Ref.~\cite{JETP_Lett_review} we take into account the spin subband splitting $\Delta_Z$ due to the Zeeman effect. For the DMS based ratchet structures the exchange enhanced Zeeman splitting is given by~\cite{Fur88,DMS2PRB12}
%}

%We assume the Zeeman splitting $\Delta_Z$ to be much smaller than the Fermi energy and take the splitting into account in the argument of oscillating terms only.

%end replaced************************

%\old{
%In comparison with Ref.~\cite{JETP_Lett_review} we take into
%account the Zeeman splitting $\Delta_Z$ in the studied DMS-based
%ratchets 
%\old{is}
%\VB{do we need "`is"'} 
%given by}
%\sdg{For the DMS based ratchet structures at low temperatures and high magnetic fields 
%we need in contrast to Ref.~\cite{JETP_Lett_review}, 
%to take the exchange enhanced Zeeman splitting $\Delta_Z$ into account.
%The latter is given by~\cite{Fur88,DMS2PRB12}}
%
\begin{equation}
\label{deltaZ} \Delta_Z = g^{*}\mu_\text{B} B + \bar{x} S_0 N_0
\alpha_e \, {\cal B}_{5/2} \left({5\mu_\text{B} g^{*}_\text{Mn} B
\over 2 k_{\rm B} (T_\text{Mn}+T_0)}\right)
\end{equation}
%
%\DW{comment: $\overline{x} = ?$}
where $g^{*}$ is the electron  Land\'{e} factor in the absence
of magnetic impurities, $\mu_{\rm B}$ is Bohr's magneton,
$g^*_\text{Mn} = 2$ is the Mn $g$-factor,
$T_\text{Mn}$ is the Mn-spin system temperature,
$S_0$ and $T_0$ account for the Mn-Mn
antiferromagnetic interaction,
${\cal B}_{5/2}(\xi)$ is the modified Brillouin function, and
$N_0 \alpha_e$ is the exchange integral.
Equations~\eqref{j},~\eqref{jy} predict that the ratchet current components 
$j_x$ and $j_y \propto j_x B_z$ have opposite parity upon magnetic field reversal
due to the photocurrent deflection by the Lorentz force~\cite{DantscherCR}.

The ratchet current becomes an oscillating function of the magnetic field
as the Landau levels move through the Fermi level and the photocurrent $1/B$-oscillation period is 
the same as for SdH oscillations.
%The theory of the magnetic ratchet effect developed in Ref.~\cite{JETP_Lett_review} shows that the ratchet current in a quantized magnetic field arises due to heating-induced corrections to the conductivity rather than a variation of the electron density. 
%Indeed, now the local electron gas heating drastically changes the conductivity in the areas heated by the near field effect. 
%This occurs because the conductivity tensor components are highly sensitive to the electron temperature variations, given by the Dingle factor~\cite{Hamaguchi}. 
%\old{DW remove: Moreover, the conductivity becomes an oscillating function of 
%the magnetic field due to periodic crossing of the Fermi level by Landau levels.}
%As a result, 
%the amplitude of the ratchet current in the magnetic field may exceed the zero-field current $j_0$ by about \sdg{two orders of }magnitude~\cite{JETP_Lett_review}.
%
%
%
%
Equation~(\ref{j}) explains the reason for
%of the observed
%sign-alternating 
the ratchet current oscillations with 
giantly enhanced magnitude as compared to the
ratchet current at zero magnetic field.
The crucial issue is that 
the ratchet current in a quantized magnetic field arises due to heating-induced corrections to the conductivity rather than a variation of the electron density, and
the conductivity in this regime is 
extremely sensitive to electron temperature variations.
%\VB{"`electron conductivity"'}
Indeed, even a weak change of the electron temperature near the Dingle temperature results in exponentially strong changes of the conductivity~\cite{Hamaguchi}. 
Therefore, the sensitivity of the conductivity to electron gas heating 
%increases strongly in quantizing magnetic fields, which 
results in a  huge enhancement of the ratchet current.
The  ratio $j(B)/j_0$ 
%ratchet current 
is governed by a factor 
${(\varepsilon_\text{F}/k_BT)(\varepsilon_\text{F}/\hbar\omega_c) \gg 1}$  
and can reach about two orders of magnitude~\cite{JETP_Lett_review}. 
%compared to the zero-field value $j_0$. 
It should be stressed that this enhancement occurs
 %and in a full agreement with experiment, 
%***SdH
%the ratchet current can be drastically enhanced compared to that at  $B=0$ 
%even 
in moderate fields where the resistance exhibits weak SdH oscillations.
%
%Moreover, the coordinate dependent
%change of the conductivity 
%caused by the inhomogeneous electron gas heating in the near field,
%explains the multi-reversal of the ratchet current direction with $B$. 
%
Heating damps the conductivity oscillations, therefore, depending on the sign of the correction, heating can either increase or decrease the conductivity because ($\partial \sigma/\partial T$) changes sign as a function of $B$.
As a result, the heating-induced ratchet current is an oscillating function of the field with a zero mean value.

\begin{figure}
        \includegraphics[width=\linewidth]{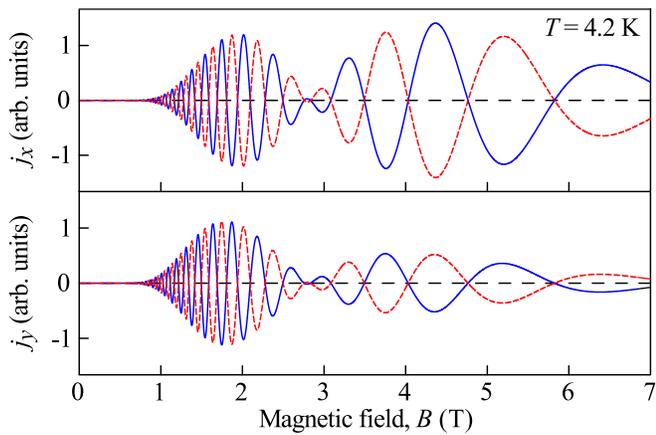}
        \caption{Photocurrents $j_x$ and $j_y$  calculated after Eqs.~\eqref{j},~\eqref{jy} for
        (Cd,Mn)Te QW sample at $T = 4.2$~K. The structure
         parameters are relevant to the experiment presented in Fig.~\ref{jxjy4K}:   $\varepsilon_{\rm F} =13$~meV,
        $\mu=0.95 \times 10^4~\mbox{cm}^2/(\mbox{Vs})$ and $\bar{x}= 0.015$.
        For other parameters we used the literature values  for \textit{n}-(Cd,Mn)Te:
$g^* = - 1.64$, $N_0 \alpha_e = 220$~meV, see Ref.~\cite{Fur88}, and $m = 0.1 \, m_0$, see Ref.~\cite{Dremin2005}. For the calculations we also used
$T_\text{Mn}=T$ and literature data of $T_0 = 0.8$~K and $S_0= 2.2$
for 
%the effective average Mn concentration 
$\bar{x} = 0.015$,
see Ref.~\cite{Ossau1993}. 
Solid and dashed curves show the results for two asymmetries of
equal magnitudes but opposite signs of $\Xi$. The oscillation amplitude is much higher than the ratchet current $j_0$ at zero magnetic field (the latter is not seen at the chosen scale).       }
        \label{Theoryjxjy}
    \end{figure}

The calculated magnetic field dependence of the ratchet currents
$j_x$ and $j_y$ are shown in Fig.~\ref{Theoryjxjy}.
The results are presented for two 
%\old{asymmetries of the electrostatic potential of} 
gate voltage combinations with equal magnitude but opposite sign of $\Xi$.
Equations~\eqref{j_zero_B}-(\ref{jy})
%Equations~(\ref{j}) and result of calculations 
reveal the most important features of the
magnetic ratchet effect:
the current is proportional to $|E_0|^2$,
%with $E_0$ being the radiation's electric field strength,
oscillates around zero with the same period as the longitudinal magnetoresistance,  
%\old{the photocurrent oscillations are sign-alternating} \fap{DW wants to remove and not replace},
%\VB{is term "period" correct for the B-dependece ? or to add something like "in 1/B scale"}
the amplitude of the current oscillations is immensely larger than the ratchet current at zero magnetic field,
and the sign of the photocurrent reverses by changing the sign of $dV/ dx$.
%
%\DW{DW comment to the parity: not obvious }\sdg{Lenia pleace add few words on additional proportionality to B for jy}
%
All these features have been observed experimentally,
see Figs.~\ref{jxjy4K}-\ref{CdTe}, \ref{asymmetry}, and \ref{temperatur}.

\begin{figure}
        \includegraphics[width=\linewidth]{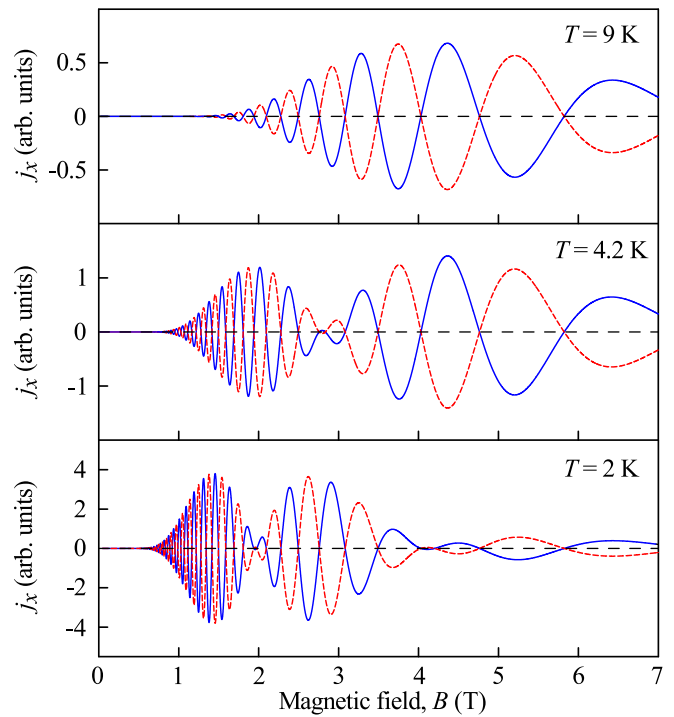}
        \caption{Photocurrent $j_x$ calculated for
        (Cd,Mn)Te QW sample at three different temperatures. Curves are obtained for
							the same values of $\varepsilon_{\rm F}$, $\mu$, $\bar{x}$, $g$, $N_0\alpha_e$, $m$, $T_{\rm Mn}$, $T_0$ and $S_0$ as used in Fig. 14. Solid and dashed curves show, similarly to Fig.~\ref{Theoryjxjy}, the results for two opposite lateral asymmetries.
				         %parameters relevant to the corresponding experiment,
        %see Fig.~\ref{jxjy4K}:  $E_{\rm F} =13$~meV,
        %$\mu=0.95 \times 10^4~\mbox{cm}^2/(\mbox{Vs})$ and $\bar{x}= 0.015$.
        %For other parameters we used literature values  for \textit{n}-(Cd,Mn)Te:
%$g = - 1.64$, $N_0 \alpha_e = 220$~meV, see Ref. [\protect
%\onlinecite{Fur88}], and $m = 0.1 \, m_0$, see Ref.
%[\onlinecite{Dremin2005}]. For the calculations we also used
%$T_{Mn}=T$ and literature data of $T_0 = 0.8$~K, and $S_0= 2.2$
%for the effective average concentration of Mn $\bar{x} = 0.015$,
%see Ref.~\cite{Ossau1993}. 
%Solid and dashed curves show the
%results for two lateral asymmetries with equal magnitude but opposite sign of $dV / dx$. The amplitude of oscillations is much
%higher than the ratchet current $j_0$ at zero field ,
%not resolved on the current scale.
}
        \label{TheoryTemp}
    \end{figure}

Equations~(\ref{j})-\eqref{delta} also describe the more complicated oscillating
ratchet current behavior observed in the DMS structures at low
temperatures, see Figs.~\ref{temperatur} and \ref{TheoryTemp} for
experiment and theory, respectively. Indeed, due to the
temperature dependence of the Brillouin function at low
temperatures  the Zeeman splitting becomes comparable to the
Landau levels separation. Calculated curves for three values of
the lattice temperatures and, consequently, different
contributions of the Zeeman effect, are shown in
Fig.~\ref{TheoryTemp}. 
%*****SdH
The calculations show, as in experiment, beats in the oscillations of $j_x(B)$.
%``beating'' effect as a function of $\bm B$. 
They stem from the interplay of Zeeman and Landau splittings
and become more pronounced with temperature decreasing.
%increase with decreasing temperature.
%\old{This explains the experimental fact showing
%that, alike the oscillations of the longitudinal resistance, the
%photocurrent oscillations at high magnetic fields and low
%temperatures {\color{red}deviate from the SdH effect behavior,} 
%see discussion in Sec.~\ref{osci}.}
%
%\VB{?? bei high magnetic fields}
Comparison of  Figs.~\ref{temperatur} and
\ref{TheoryTemp} demonstrates that Eq.~(\ref{j}) 
describes the qualitative behavior of the photocurrent quite well. 
We note that the beats in $j_{x,y}(B)$ are  substantially influenced by the Mn spin system temperature,
 which, because of electron gas heating, can be higher than the lattice one~\cite{Kel02}. 
This increase  due to the strong dependence of the Brillouin function on $T_\text{Mn}$, 
%\old{DW remove: of Mn spin system temperature} 
results in a reduced exchange part of $\Delta_Z$ in Eq.~(\ref{deltaZ}). 
This shifts the onset of the exchange enhanced 
Zeeman splitting to higher magnetic fields.
Allowance for this effect should further improve the agreement between experimental data and calculations.

%\FloatBarrier

\section{Summary} \label{summary}

To summarize, we have experimentally demonstrated and
theoretically explained the magnetic quantum ratchet effect,
 i.e., the ratchet effect in quantizing magnetic fields. To generate 
the ratchet current we use an asymmetric interdigitated top 
gate superlattice on top of the investigated quantum wells. 
 Electric currents driven by
terahertz electric field exhibit sign-changing
magneto-oscillations with an amplitude giantly enhanced as 
compared to the photocurrent at zero magnetic field. The current
amplitude and direction can be controllably changed by the
variation of voltages applied to individual gate sublattices.
The effect is observed in structures with various QW compositions
and superlattice parameters. The photocurrent generation mechanism can be
well described in terms of semiclassical theory of magnetic
ratchet effects. 
%excited by an $ac$ electric field. 
The observed effect is driven by the periodic modulation
of the electron temperature caused by near field
diffraction. The theory of the Seebeck ratchet effect in
the presence of quantizing magnetic field shows that the ratchet current follows the oscillations of the longitudinal resistance. In the
DMS structures at low temperatures %and high magnetic field 
the investigated
effect is strongly influenced by the exchange enhanced Zeeman
splitting caused by the exchange interaction of electrons with Mn$^{2+}$ ions.

%\fap{in REF [38]  the phys.rev. b is not 85 it should be 86.
% in REF [ 50] the sentences are somehow double included}

Finally we note, that while here
we have dealt with 
%in experiments and theory the effect of 
a homogeneous
magnetic field, the observed ratchet photocurrent in
superlattices made of hard magnetic material (Dy) opens a possibility
to study the ratchet effect driven by
an inhomogeneous periodic magnetic field recently suggested in
Ref.~\cite{JETP_Lett_review,Budkin_Golub}. 
This kind of experiments would apply
remnant magnetization of Dysprosium together with the enhanced
magnetic properties of DMS QWs and is a future task.

\section{Acknowledgments} \label{acknow}

We thank I. A. Dmitriev and D. R. Yakovlev for helpful discussions.
The support from the DFG priority program SFB~689,
the Volkswagen Stiftung Program and RFBR (projects 15-02-02989,
%Popov
16-02-01037,
16-02-00375, and 17-52-53063)
%LG
is gratefully acknowledged. The work of DVF was
supported by the Grant from the President of the Russian
Federation (No. MK-5447.2016.2). The research in Poland was
partially supported by the National Science Centre (Poland)
through Grant No. DEC-2012/06/A/ST3/00247 and by the Foundation
for Polish Science through the IRA Programme financed by EU within
SG OP Programme.


\begin{thebibliography}{99}




%in the absence of an average macroscopic force resulting in the ratchet effect~\cite{x}.
%~\cite{prost,reimann,applphys,hanggi,Denisovhanggi}.
%
%If the directed transport in an asymmetric
%periodic system is induced by electro-magnetic radiation, it is usually
%referred to as photogalvanic (or sometimes photovoltaic)
%effect~\cite{BelStu,OurPRL,OurPRL2,10,book,kotthaus,1,2,3,regensburg,samuelson}, particularly
%if breaking of spatial inversion symmetry is related to the microscopic
%structure of the system.
%
\bibitem{prost} F. J\"{u}licher, A. Ajdari, and J. Prost, Rev. Mod. Phys. {\bf 69}, 1269 (1997).
%
\bibitem{reimann} P. Reimann,
%\textit{Brownian motors: Noisy transport far from equilibrium}
Phys. Rep. {\bf 361}, 57 (2002).
%\bibitem{reimann} P. Reimann, Phys. Rep. {\bf 361}, 57 (2002), ``Brownian motors: Noisy transport far from equilibrium''.
%
\bibitem{applphys} H. Linke (ed.), \emph{Ratchets and Brownian Motors: Basics, Experiments and
Applications}, special issue Appl. Phys. A: Mater. Sci. Process. A
{\bf 75}, 167 (2002).
%\bibitem{special_issue} H. Linke, Applied Physics A: Materials Science Processing {\bf 75}, 167 (2002).

\bibitem{hanggi} P. H\"{a}nggi and F. Marchesoni, Rev. Mod. Phys. {\bf 81}, 387 (2009).

\bibitem{Denisovhanggi}S. Denisov, S. Flach, and P. H\"{a}nggi, Phys. Rep. {\bf 538}, 77 (2014).

%Ratchet effects whose prerequisites are simultaneous breaking of both thermal equilibrium and spatial inversion symmetry can be realized in a great variety of forms and in particular, as electric transport in semiconductor systems~\cite{buttiker,buttiker2,grifoni,kotthaus,samuelson,Chepel,Chepelianskii,Kvon,Olbrich_PRL_09,Kannan,Olbrich_PRB_11,}.

\bibitem{buttiker} M. B\"uttiker, Z. Phys. B {\bf 68}, 161 (1987).
%
\bibitem{buttiker2} Ya.~M. Blanter and M. B\"uttiker, Phys. Rev. Lett. {\bf 81}, 4040 (1998).
%
\bibitem{grifoni} P. Reimann, M. Grifoni, and P. H\"anggi, Phys. Rev. Lett. {\bf 79}, 10 (1997).

\bibitem{kotthaus} A. Lorke \emph{et al}., Physica B {\bf 249}, 312 (1998).
%
\bibitem{samuelson} A.~M. Song ~\emph{et al}.,
%P. Omling, L. Samuelson, W. Seifert, and I. Shorubalko, H. Zirath,
Appl. Phys. Lett. {\bf 79}, 1357 (2001).

\bibitem{Chepel} A. D. Chepelianskii, M. V. Entin, L. I. Magarill, and D. L.
Shepelyansky, Eur. Phys. J. \textbf{56}, 323 (2007).

\bibitem{Chepelianskii}  A.~D. Chepelianskii, M.~V. Entin, L.~I. Magarill, and D.~L. Chepelyansky, Eur. Phys. J. B {\bf 56},
323 (2007); Physica E (Amsterdam) {\bf 40}, 1264 (2008).

\bibitem{Kvon} S. Sassine, Yu. Krupko, J.-C. Portal, Z. D. Kvon, R. Murali, K. P. Martin, G. Hill, and A. D. Wieck,
Phys. Rev. B \textbf{78}, 045431 (2008).

\bibitem{Olbrich_PRL_09} P. Olbrich, E. L. Ivchenko, R. Ravash, T. Feil, S. N. Danilov,
J. Allerdings, D. Weiss, D. Schuh, W. Wegscheider, and S. D. Ganichev,
%\textit{Ratchet effects induced by terahertz radiation in heterostructures with a lateral periodic potential}
{Phys. Rev. Lett.} \textbf{103}, 090603 (2009).

\bibitem{Review_JETP_Lett} E. L. Ivchenko and S. D. Ganichev,
%\textit{Ratchet effects in quantum wells with a lateral superlattice}
{Pisma v ZheTF} \textbf{93}, 752 (2011) [{JETP Lett.} \textbf{93}, 673 (2011)].

\bibitem{Nalitov}  A. V. Nalitov, L. E. Golub, and E. L. Ivchenko, Phys. Rev. B \textbf{86}, 115301 (2012).
%\textit{Ratchet effects in two-dimensional systems with a lateral periodic potential}


\bibitem{PopovAPL2013} V. V. Popov, Appl. Phys. Lett. \textbf{102}, 253504 (2013).
%\textit{Terahertz rectification by periodic two-dimensional electron plasma}

%\bibitem{Koniakhin} S. V. Koniakhin, Eur. Phys. J. B \textbf{87}, 216 (2014).
\bibitem{Koniakhin2014} S. V. Koniakhin,
%\textit{Ratchet effect in graphene with trigonal clusters}
Eur. Phys. J. B  \textbf{87}, 216 (2014).

\bibitem{Rozhansky2015} I. V. Rozhansky, V. Yu. Kachorovskii, and M. S. Shur,
%\textit{Helicity-driven ratchet effect enhanced by plasmons}
%arXiv:1411.7436v1
Phys. Rev. Lett. \textbf{114}, 246601 (2015).

\bibitem{PopovIvchenko} V. V. Popov, D. V. Fateev, E. L. Ivchenko, and S. D. Ganichev,
Phys. Rev. B \textbf{91}, 235436 (2015).
%\\\textit{Noncentrosymmetric plasmon modes and giant terahertz photocurrent in a
%two-dimensional plasmonic crystal}\\

\bibitem{Olbrich_PRB_11} P. Olbrich, J. Karch, E. L. Ivchenko, J. Kamann, B. M\"{a}rz, M. Fehrenbacher, D. Weiss, and S. D. Ganichev,
%\textit{Classical ratchet effects in heterostructures with a lateral periodic potential}
{Phys. Rev. B} \textbf{83}, 165320 (2011).

\bibitem{Kannan} E. S. Kannan, I. Bisotto, J.-C. Portal, T. J. Beck, and L. Jalabert, Appl.
Phys. Lett. \textbf{101}, 143504 (2012).

%can be efficiently excited
%in semiconductor quantum wells (QW) and graphene with a lateral superlattice structures~\cite{x}.
%~\cite{Olbrich_PRL_09,Olbrich_PRB_11,Review_JETP_Lett,popovAPL,otsuji,otsuji2,otsuji3,faltermeier15},

\bibitem{ratchet_graphene} P. Olbrich, J. Kamann, M. K\"{o}nig, J. Munzert, L. Tutsch, J. Eroms, D. Weiss,
M.-H. Liu, L. E. Golub, E. L. Ivchenko, V. V. Popov, D. V. Fateev, K. V. Mashinsky, F. Fromm, Th. Seyller, and S. D. Ganichev, Phys. Rev. B \textbf{93}, 075422 (2016).

\bibitem{Drexler13} C.~Drexler,  S.~A.~Tarasenko, P.~Olbrich, J.~Karch,
M.~Hirmer, F. M\"{u}ller, M.~Gmitra, J. Fabian,
R.~Yakimova, S.~Lara-Avila, S.~Kubatkin, and S.~D.~Ganichev,
%Magnetic quantum ratchet effect in graphene,
Nat. Nanotechn. \textbf{8}, 104 (2013).

\bibitem{otsuji} T. Watanabe, S. A. Boubanga-Tombet, Y. Tanimoto, D. Fateev, V. Popov, D. Coquillat, W. Knap, Y. M. Meziani, Yuye Wang, H. Minamide, H. Ito, and T. Otsuji,
%\textit{InP- and GaAs-based plasmonic high-electron-mobility transistors for room-temperature ultrahigh-sensitive terahertz sensing and imaging}
IEEE Sensors J. {\bf 3}, 89 (2013).
\bibitem{det2} Y. Kurita, G. Ducournau, D. Coquillat, A. Satou1, K. Kobayashi,
S. Boubanga Tombet, Y. M. Meziani, V. V. Popov, W. Knap,
T. Suemitsu, and T. Otsuji, Appl. Phys. Lett. \textbf{104}, 251114
(2014).
\bibitem{otsuji2} S. A. Boubanga-Tombet, Y. Tanimoto, A. Satou, T. Suemitsu, Y. Wang, H. Minamide, H. Ito,
D. V. Fateev, V. V. Popov, and T. Otsuji,
%\textit{Current driven detection of terahertz radiation in dual-grating-gate plasmonic detector}
Appl. Phys. Lett. {\bf 104}, 262104 (2014).
\bibitem{Popov_Otsuji_Knap} V. V. Popov, D. V. Fateev, T. Otsuji, Y. M. Meziani, D. Coquillat,
and W. Knap, Appl. Phys. Lett. \textbf{99}, 243504 (2011).


\bibitem{Otsuji_Ganichev}  P. Faltermeier, P. Olbrich, W. Probst, L. Schell, T. Watanabe,
S. A. Boubanga-Tombet, T. Otsuji, and S. D. Ganichev,
J. Appl. Phys. \textbf{118}, 084301 (2015).
%\bibitem{faltermeier15} P. Faltermeier, P. Olbrich,  W. Probst, L. Schell,
%T. Watanabe, S. A. Boubanga-Tombet, T. Otsuji, and
%and S. D.\,Ganichev
%\\\textit{Helicity sensitive terahertz radiation detection by
%dual-grating-gate high electron mobility transistors}\\
%J. Appl. Phys. \textbf{118}, 084301 (2015).

\bibitem{helicitydetector} S. D. Ganichev, W. Weber, J. Kiermaier, S. N. Danilov, D. Schuh, W. Wegscheider, Ch. Gerl, D. Bougeard, G. Abstreiter and W. Prettl,
% All-electric detectors of the polarization state of terahertz laser radiation,
J. Appl. Physics \textbf{103}, 114504 (2008).

\bibitem{ellipticitydetector}
%Fast  detector of  the ellipticity of infrared and terahertz radiation based on HgTe quantum well structures,
S. N. Danilov, B. Wittmann, P. Olbrich, W. Eder, W. Prettl, L. E.
Golub, E. V.Beregulin, Z. D. Kvon, N. N. Mikhailov, S. A.
Dvoretsky, V. A. Shalygin, N. Q. Vinh, A. F. G. van der Meer, B.
Murdin, and S. D. Ganichev, J. Appl. Phys. \textbf{105}, 013106
(2009).

\bibitem{ellipticitydetector2}
S. Dvoretsky, N. Mikhailov, Y. Sidorov, V. Shvets, S. Danilov, B.
Wittman, and S. Ganichev,
%\textit{Growth of HgTe quantum wells for IR to THz Detectors}
J. Electron. Mat. \textbf{39},  918 (2010).

\bibitem{JETP_Lett_review} G. V. Budkin, L. E. Golub, E. L. Ivchenko, and S. D. Ganichev, JETP Lett. \textbf{104}, 649 (2016).

%
%%material
\bibitem{Crooker}  S. A.~Crooker,
D.~A.~Tulchinsky, J.~Levy, D.~D.~Awschalom, R.~Garcia, and
N.~Samarth, Phys. Rev. Lett. \textbf{75}, 505 (1995).

\bibitem{Egues} J. C.~Egues and J. W.~Wilkins, Phys. Rev. \textbf{58}, R16012 (1998).

\bibitem{Jaroszynski2002} J. Jaroszynski,
T. Andrearczyk, G.~Karczewski, J.~Wr\'{o}bel, T.~Wojtowicz,
E.~Papis, E.~Kaminska, A.~Piotrowska, D.~Popovic, and T.~Dietl,
Phys. Rev. Lett. \textbf{89}, 266802 (2002).

\bibitem{DMSPRL09} S. D.~Ganichev, S. A. Tarasenko, V. V. Bel'kov, P. Olbrich, W. Eder, D. R.~Yakovlev, V. Kolkovsky, W. Zaleszczyk,
G.~Karczewski, T. Wojtowicz, and D. Weiss,
%\textit{Spin currents in diluted magnetic semiconductors}
Phys. Rev. Lett. \textbf{102}, 156602 (2009).

\bibitem{DMS2PRB12}  P. Olbrich, C. Zoth, P. Lutz, C. Drexler, V. V. Bel'kov, Ya.
V. Terent'ev, S. A. Tarasenko, A. N. Semenov, S. V. Ivanov, D. R.
Yakovlev, T. Wojtowicz, U. Wurstbauer, D. Schuh, and S.
D.\,Ganichev,
%\\\textit{Spin-polarized electric currents in diluted magnetic
%semiconductor heterostructures  induced  by terahertz and microwave radiation}\\
Phys. Rev. B \textbf{86}, 085310 (2012).


\bibitem{Kneip2006} M. K. Kneip, D.~R.~Yakovlev,  M.~Bayer, G.~Karczewski, T.~Wojtowicz, and J.~Kossut,
Appl. Phys. Lett. \textbf{88}, 152105 (2006).
%Engineering of spin-lattice relaxation dynamics by digital growth of diluted magnetic semiconductor CdMnTe


\bibitem{Gaj79} J. A. Gaj, R. Planel, and
G. Fishman, Solid State Commun. \textbf{29}, 435 (1979).
%Relation of magneto-optical properties of free excitons to spin alignment of Mn2+ ions in Cd1−xMnxTe

\bibitem{Fur88} J. K. Furdyna, J. Appl. Phys. \textbf{64}, R29 (1988).
%Diluted magnetic semiconductors

\bibitem{Dietl} T.~Dietl, in \textit{Handbook on Semiconductors}, vol.\,3b,
Ed. T.S. Moss (North-Holland, Amsterdam, 1994).

\bibitem{DMS2010} \textit{Introduction to the Physics of Diluted Magnetic Semiconductors},
Eds. J. Kossut and J.\,A. Gaj  (Springer, Berlin 2010).



\bibitem{staab2015} M. Staab, M. Matuschek, P. Pereyra, M. Utz, D. Schuh, D. Bougeard,
R. R. Gerhardts, and D. Weiss,
%Commensurability oscillations in a lateral superlattice with
%broken inversion symmetry.
New J. Phys. \textbf{17}, 043035 (2015).

\bibitem{Budkin_Golub} %Orbital magnetic ratchet effect,
G.V. Budkin, and L.E. Golub, Phys. Rev. B \textbf{90}, 125316 (2014).


\bibitem{hallgraphene} J. Karch, P. Olbrich, M.~Schmalzbauer, C. Zoth, C.~Brinsteiner,
M. Fehrenbacher,  U.~Wurstbauer, M.\,M.~Glazov, S. A. Tarasenko,
E. L. Ivchenko, D. Weiss, J.~Eroms, and S. D.~Ganichev,
%\\\textit{Dynamic Hall effect driven by circularly polarized light in a graphene layer},\\
Phys. Rev. Lett.  \textbf{97}, 227402 (2010).

\bibitem{Kvon2008}  Z.\,D. Kvon, S.\,N. Danilov, N.\,N. Mikhailov, S.\,A. Dvoretsky, and S.\,D. Ganichev,
%Cyclotron resonance photoconductivity of a two-dimensional electron gas in HgTe quantum wells,
Physica E \textbf{40}, 1885 (2008).




\bibitem{Ziemann2000}
%S.D.~Ganichev,
%%\\{\it The Infrared Spin-Galvanic Effect in Semiconductor Quantum Wells},\\
%Physica E {\bf 20}, 419 (2004).
E. Ziemann, S. D. Ganichev, I. N. Yassievich, V. I. Perel, and W.
Prettl, J. Appl. Phys. \textbf{87}, 3843 (2000).

\bibitem{footnoteJD} {Note that in $x$-direction
the polarization independent contribution $J^D_x$ dominates the
photocurrent for all studied 
%(Cd,Mn)Te-based  and CdTe-based
samples and almost all combinations of gate voltages applied to the
individual subgratings of DGG.}

\bibitem{footnoteimperfections} {Fabrication of such a large-size superlattices is a challenging
task. Therefore, some of our structures have minor imperfections,
as confirmed by optical microscope images. The imperfections occur
on samples with occasional inhomogeneities of the PMMA resist.
Due to that the electron beam lithography was unable to write
the exact width of the stripes as it was intended and, therefore,
ended up with an overlap of the Au fingers. This leads to  short
circuits, as well as to complete destruction of few gate fingers
by removing the additional Au material in the finger spacings. In
some of our structures the connection between two gate stripes
also showed up when surface impurities were covered with Au, which
were not supposed to be part of the gate fingers.}

\bibitem{Belkov2008} V. V. Bel'kov and S. D. Ganichev,
% \\\textit{Magneto-gyrotropic effects in semiconductor quantum wells}\\
Semicond. Sci. Technol. \textbf{23}, 114003 (2008).











\bibitem{footnotescan}
%Note that for large shifts of the beam, for which laser spot approach
%sample edges the signal starts to grow again.
%This result is attributed to the generation of
%the edge photocurrents reported for graphene in Refs.~\cite{Karch_PRL_2011,Glazov2014}
%and also observed for semiconductor QWs~\cite{NTTI2014}.
The edge photocurrent is caused by the asymmetric scattering of carriers 
driven back and force to the edge by the radiation electric field~\cite{Karch_PRL_2011,Glazov2014}.


\bibitem{Karch_PRL_2011}    J. Karch, C. Drexler, P. Olbrich, M. Fehrenbacher, M. Hirmer, M. M. Glazov,
S. A. Tarasenko, E. L. Ivchenko, B. Birkner, J. Eroms, D. Weiss, R. Yakimova, S. Lara-Avila, S. Kubatkin,
M. Ostler, T. Seyller, and S. D. Ganichev,
%\textit{Terahertz radiation driven chiral edge currents in graphene}
Phys. Rev. Lett. \textbf{107}, 276601 (2011).

\bibitem{Glazov2014}  M. M.\,Glazov and S. D.\,Ganichev,
%\textit{High frequency electric field induced nonlinear effects in graphene}
Physics Reports \textbf{535},  101 (2014).
%\bibitem{Glazov2014}  M.M.\,Glazov and S.D.\,Ganichev,
%\textit{High frequency electric field induced nonlinear effects in graphene}, Physics Reports \textbf{535},  101 (2014).

\bibitem{NTTI2014}
%Photogalvanic probing of helical edge channels in 2D HgTe topological insulators}
K.-M.\,Dantscher,  D.\,A.\,Kozlov,
M.\,T.\,Scherr, S.\,Gebert, J.\,B\"arenf\"anger,
M.\,V.\,Durnev, S.\,A.\,Tarasenko,
V.\,V.\,Bel'kov,  N.\,N.\,Mikhailov,
S.\,A.\,Dvoretsky, Z.\,D.\,Kvon,
D.\,Weiss, and S.\,D.\,Ganichev, arXiv cond-mat: 1612.08854 (2016).

\bibitem{Fateev2010} D. V. Fateev, V. V. Popov, and M. S. Shur,
Semicond. \textbf{44}, 1406 (2010).
%\bibitem{Fateev2010} D.V. Fateev, V.V. Popov, and M.S. Shur, Semiconductors \textbf{44}, 1406 (2010).




%\bibitem{Shangina2011}
%E.~L. Shangina, K.~V. Smirnov, D.~V. Morozov, V.~V. Kovalyuk, 
%G.~N. Goltsman, A.~A. Verevkin, A.~I. Toropov and P. Mauskopf, 
%Semicond. Sci. Technol. \textbf{26}, 025013 (2011).

\bibitem{DasSarma1990} T. Kawamura, S. Das Sarma, R. Jalabert, and J. K. Jain, Phys. Rev. B \textbf{42}, 5407 (1990).

\bibitem{DantscherCR} K.-M.\,Dantscher, D. A. Kozlov, P. Olbrich, C. Zoth,
P.\,Faltermeier, M.\,Lindner, G. V.\,Budkin, S. A. Tarasenko, V.
V.\,Bel'kov, Z. D.\,Kvon, N. N. Mikhailov, S. A. Dvoretsky,
D.\,Weiss, B. Jenichen, and S. D.\,Ganichev,
%\\\textit{Cyclotron Resonance Assisted Photocurrents in Surface States of a 3D Topological Insulator Based on a Strained High Mobility HgTe Film}\\
 Phys. Rev. B \textbf{92}, 165314 (2015).


\bibitem{Hamaguchi} C. Hamaguchi, \textit{Basic Semiconductor Physics}
(Springer 2014).

\bibitem{Dremin2005} A. A. Dremin, D. R. Yakovlev, A. A. Sirenko, S. I. Gubarev, O. P. Shabelsky, A. Waag, and M. Bayer, Phys. Rev. B
 \textbf{72}, 195337 (2005).
%Electron cyclotron mass in undoped CdTe/CdMnTe quantum wells

\bibitem{Ossau1993} W. J. Ossau and B. Kuhn-Heinrich,
%Dimensional dependence of antiferromagnetism in
%diluted magnetic semiconductor heterostructures
Physica B \textbf{184}, 422 (1993).








\bibitem{Kel02} D. Keller,
% et al., 
D.~R.~Yakovlev, B.~K\"{o}nig, W.~Ossau,
Th.~Gruber, A.~Waag, L.~W.~Molenkamp, and A.~V.~Scherbakov, 
Phys. Rev. B \textbf{65}, 035313 (2002). 




\end{thebibliography}
\end{document}